\documentclass[preprintnumbers,showkeys,amsmath,amssymb]{revtex4}
\usepackage{graphicx}
\usepackage{dcolumn}
\usepackage{bm}
\usepackage{epstopdf}
\usepackage{amsmath}
\usepackage{multirow}
\usepackage{float}

\usepackage{soul,color,xcolor}
\soulregister\upcite7
\soulregister\citep7
\soulregister\citet7
\soulregister\ref7
\soulregister\pageref7
\soulregister\and7
\soulregister\Name7

\begin{document}

\preprint{}

\title{Complete hyperentangled Bell states analysis for polarization-spatial-time-bin degrees of freedom with unity fidelity}

\author{Xin-Jie Zhou$^{1}$, Wen-Qiang Liu$^{1,2}$, Yan-Bei Zheng$^{1}$, Hai-Rui Wei$^{1}$,\footnote{Corresponding author: hrwei@ustb.edu.cn} and Fang-Fang Du$^{3}$}
\address{
 $^1$School of Mathematics and Physics, University of Science and Technology Beijing, Beijing 100083, China\\
 $^2$Center for Quantum Technology Research and Key Laboratory of Advanced Optoelectronic Quantum Architecture and Measurements (MOE), School of Physics, Beijing Institute of Technology, Beijing 100081, China\\
 $^3$ Science and Technology on Electronic Test and Measurement Laboratory, North University of China, Taiyuan 030051, China}

\begin{abstract}
Hyperentangled states can outperform their classical counterparts on solving certain tasks. Here we present a simplified scheme for completely distinguishing two-photon hyperentangled Bell states in polarization, spatial, and time-bin degrees of freedom (DOFs).  Unity fidelity can be achieved in principle without strong couple limitation between photon and quantum dot (QD), and the incomplete and imperfect QD-cavity interactions are prevented by single-photon detectors. In addition, auxiliary photons or DOFs are not required in our scheme. The necessary linear optical elements are fewer than the parity-check-based one.
\end{abstract}



\maketitle

\newcommand{\upcite}[1]{\textsuperscript{\textsuperscript{\cite{#1}}}}

\section{Introduction}\label{sec1}


Entanglement, a key quantum resource, plays a critical role in many important applications in quantum information processing (QIP),\upcite{CITE1} including  quantum key distribution,\upcite{distribution1,distribution2,quantum-key-distribution1} quantum dense coding,\upcite{dense1,dense2,Advances-in-Quantum-Dense-Coding} quantum teleportation,\upcite{teleportation,teleportation1,teleportation2} quantum secret sharing,\upcite{share,share2} quantum secure direct
communication,\upcite{direct1,direct2,direct3,direct4,direct5} quantum networks,\upcite{quantum-router} and one-way quantum computation.\upcite{one-way1,one-way2}
The entanglement of particle pairs simultaneously exists in more than one degree of freedom (DOF),\upcite{multiple-DOFs2,multiple-DOFs3,multiple-DOFs4,multiple-DOFs5,multiple-DOFs6,multiple-DOFs7,multiple-DOFs1} referred to as hyperentanglement,\upcite {distribution1} and that is subject to high capacity, low loss rate, less quantum resources, and loss decoherence characters.
Hyperentanglement has gained widespread attention in recent years due to its excellent properties,\upcite{multiple-DOFs3,multiple-DOFs4,multiple-DOFs5,multiple-DOFs2,multiple-DOFs6} and it has been widely used in linear optical quantum dense coding,\upcite{multiple-DOFs7} teleportation-based quantum networking,\upcite{teleportation-based} deterministic entanglement purification,\upcite{purification1,purification2,purification3} entanglement witness,\upcite{entanglement-witness} one-way quantum computing,\upcite{one-way-qunatum-computing} quantum key distribution,\upcite{QKD-hyper} linear-optical heralded amplification,\upcite{Linear-optical-heralded-amplification} etc.
Moreover, hyperentangled states can be utilized to complete tasks which are challenged in single DOF systems, such as complete Bell states analysis (BSA) with linear optical elements.\upcite{guoguangcan,ptd}



BSA, which is defined as a distinction between the four maximally entangled Bell states, is an important step in many quantum communication protocols. Complete BSA has been realized by cross-Kerr nonlinearity,\upcite{BSA-kerr,purification1} nuclear magnetic resonance system,\upcite{NMR} atomic system,\upcite{atom} cavity quantum electrodynamics system,\upcite{QED} and quantum dot (QD).\upcite{BSA-QD,BSA-QDs,hu-QD,Cao-QD}
Unfortunately, a complete BSA is impossible by using standard linear optical methods and a single degree of shared entanglement.\upcite{not-possible}
While a BSA with linear optics can be  completely performed assisted by hyperentangled states,\upcite{guoguangcan,ptd}  single-photon entangled states,\upcite{single-photon-BSA} or nonlinear mediums.\upcite{BSA-kerr,NMR,atom,QED,BSA-QD,BSA-QDs} The complete linear-optical BSA has been demonstrated using hyperentangled states in polarization-spatial DOF, polarization-orbital-angular-momentum DOF, polarization-momentum DOF (which are not compatible with transmission through fiber-based networks),\upcite{multiple-DOFs7,multiple-DOFs3,teleportation-based} polarization-time DOF (number-resolving detectors are required),\upcite{multiple-DOFs2} and polarization-time-delay DOF.\upcite{ptd}


Tremendous progress has also been made in hyperentangled Bell states analysis (HBSA).\upcite{seven-groups,sheng,ren-QD,wang-QD,wang-error-detected-QD,Error-heralded,Cao,
polarization-time-bin,liu-two-photon-six-qubit,three-DOF-kerr-wangmeiyu,three-DOF-kerr-Zhang,Self-assisted,auxiliary-entanglement,time-bin1,time-bin2}
In 2007, Wei \emph{et al}.\upcite{seven-groups} shown that 16 hyperentangled Bell (hyper-Bell) states can not be unambiguously distinguished only using linear optics.
Later, the first complete HBSA scheme was proposed by Sheng \emph{et al}.\upcite{sheng} in 2010 via cross-Kerr nonlinearity, and another interesting HBSAs via cross-Kerr nonlinearity were proposed later.\upcite{polarization-time-bin,liu-two-photon-six-qubit,three-DOF-kerr-wangmeiyu,three-DOF-kerr-Zhang,Self-assisted}
Giant Kerr nonlinear is a challenge in experiment.
In 2012, Ren \emph{et al}.\upcite{ren-QD} and Wang \emph{et al}.\upcite{wang-QD} proposed schemes to complete HBSA schemes in polarization and spatial DOFs via QD-cavity system.
Such two schemes were further improved to the error-detected ones by Wang \emph{et al}.,\upcite {wang-error-detected-QD} Zheng \emph{et al}.,\upcite{Error-heralded} and Cao \emph{et al}.\upcite{Cao}
QDs, named as artificial atom, have been recognized as the promising candidates for QIP.\upcite{QD-cavity-gate-hybrid1,QD-cavity-gate-hybrid2,QD-cavity-gate-hybrid3,QD-cavity-gate-solid1,QD-cavity-gate-solid2,QD-cavity-gate-photon1,QD-cavity-gate-photon2,
QD-nature,QD-spin,teleportation3,QD-NC43,QD-NC44,QD-switch,QD-rotation1,QD-rotation2,QD-entangle-photon}
Nowadays, schemes for implementing quantum gates have been proposed in hybrid photon-QD systems,\upcite{QD-cavity-gate-hybrid1,QD-cavity-gate-hybrid2,QD-cavity-gate-hybrid3} solid-QD systems,\upcite{QD-cavity-gate-solid1,QD-cavity-gate-solid2} and flying photon systems,\upcite{QD-cavity-gate-photon1,QD-cavity-gate-photon2} respectively.
Quantum entanglements between a QD spin and a single photon\upcite{QD-nature} and  between two distance QD hole spins\upcite{QD-spin} have been experimentally demonstrated.
Quantum teleportation from a flying single photon to a solid-state QD spin was experimentally realized by Gao \emph{et al}.\upcite{teleportation3}
Utilizing QD-cavity combination, a QD photon-sorter,\upcite{QD-NC43,QD-NC44} a spin-photon quantum phase switch,\upcite{QD-switch} and Faraday rotation induced by a single hole spin\upcite{QD-rotation1} or an electron spin\upcite{QD-rotation2} in QDs have been experimentally realized.
The photon-QD platforms contained in above schemes usually are not unity.


In this paper, we design a scheme to completely distinguish hyper-Bell states in polarization, spatial, and time-bin DOFs assisted by single-sided QD-cavity systems.
The 64 hyper-Bell states are divided into 8 groups with the help of three QD-cavity systems in terms of spatial and polarization DOFs.
And then, the 8 hyper-Bell states in each distinct group can be distinguished from each other by the detection signatures of the single-photon detectors, see Table \ref{table1}.
Compared with the previous auxiliary-based schemes (such as auxiliary entangled state, parity-check quantum nondestructive detectors), the unity-fidelity of our scheme can be achieved with less nonlinear photon-mater interactions and linear optical elements.
Imperfect and incomplete cavity-QD interactions are all prevented by single-photon detectors, and strong couple between QDs and cavities is also not necessary in our scheme.

\section{Complete hyper-Bell states analysis for polarization-spatial-time-bin DOFs using QD-cavity systems}\label{sec2}

\subsection{Interactions between the photon and the QD in a single-sided micropillar  microcavity}\label{sec2.1}

\begin{figure} [htpb]
\begin{center}
\includegraphics[width=13 cm,angle=0]{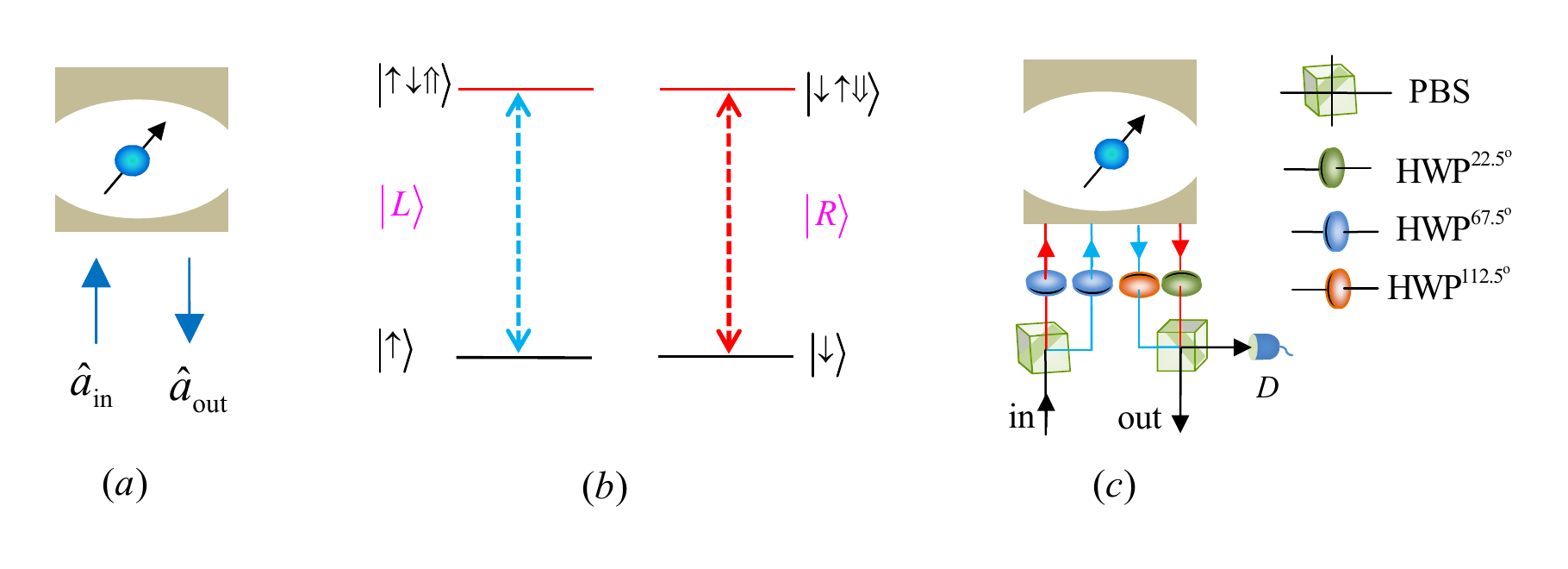}
\caption{(a) A singly charged QD inside a single-sided optical micropillar cavity. (b) Spin-dependent optical transition rules. (c) Heralded unity-fidelity emitter.
HWP$^{22.5^\circ}$, HWP$^{67.5^\circ}$, and HWP$^{112.5^\circ}$  represent half-wave plates oriented at $22.5^\circ$, $67.5^\circ$, and $112.5^\circ$, respectively.
Each PBS represents a polarization beam splitter which transmits the $R$-polarized and reflects the $L$-polarized photons, respectively. $D$ represents a single-photon detector. \label{singleQD}}
\end{center}
\end{figure}

Let us firstly review the emitter, i.e., a singly charged QD embedded in the center of a single-sided optical microcavity.\upcite{QD-basic} As shown in Figure \ref{singleQD} (a) and Figure \ref{singleQD} (b), a negatively charged exciton $X^{-}$ consisted of two electrons bound to one hole can be created by injecting an excess electron into the QD.\upcite{exciton}  According to the Pauli’s exclusion principle, $X^{-}$ conducts the spin-dependent optical transition rules,\upcite{QD-Pauli-principle} i.e., left-handed circularly polarized photon (marked by $L$) couples to $|\uparrow\rangle \rightarrow |\uparrow\downarrow\Uparrow\rangle$  and right-handed circularly polarized photon (marked by $R$) couples to $|\downarrow\rangle \rightarrow |\downarrow\uparrow\Downarrow\rangle$. Here $|\uparrow\rangle\equiv|+1/2\rangle$, $|\downarrow\rangle\equiv|-1/2\rangle$, $|\Uparrow\rangle\equiv|+3/2\rangle$, and $|\Downarrow\rangle\equiv|-3/2\rangle$.
HWP$^{22.5^\circ}$, HWP$^{67.5^\circ}$, and HWP$^{112.5^\circ}$ shown in Figure \ref{singleQD}(c), present half-wave plates oriented at $22.5^\circ$, $67.5^\circ$, and $112.5^\circ$, resulting in
\begin{eqnarray}              \label{eq5}
\begin{split}
&| R \rangle \xrightarrow{\text{HWP}^{22.5^\circ}} \frac{1}{\sqrt{2}}(| R \rangle+| L \rangle),\qquad\;
 | L \rangle \xrightarrow{\text{HWP}^{22.5^\circ}} \frac{1}{\sqrt{2}}(| R \rangle-| L \rangle),\\
&| R \rangle \xrightarrow{\text{HWP}^{67.5^\circ}} \frac{1}{\sqrt{2}}(-|R \rangle+| L \rangle),\quad\;\;\;
 | L \rangle \xrightarrow{\text{HWP}^{67.5^\circ}} \frac{1}{\sqrt{2}}(| R \rangle+| L \rangle),\\
&| R \rangle \xrightarrow{\text{HWP}^{112.5^\circ}} -\frac{1}{\sqrt{2}}(|R \rangle+| L \rangle),\quad\;\;
 | L \rangle \xrightarrow{\text{HWP}^{112.5^\circ}} \frac{1}{\sqrt{2}}(-|R \rangle+| L \rangle).
\end{split}
\end{eqnarray}


The Heisenberg equations for the cavity field operator $\hat{a}$ and the QD dipole operator $\hat{\sigma}_-$ and $\hat{\sigma}_z$ are given by\upcite{Heisenberg1,Heisenberg2}
\begin{equation}              \label{eq1}
\begin{aligned}
&\frac{{d} \hat{a}}{{~d} t} =-\left[{i}\left(\omega_{c}-\omega\right)+\frac{\kappa}{2}+\frac{\kappa_{s}}{2}\right] \hat{a}-g \hat{\sigma}_{-}-\sqrt{\kappa} \hat{a}_{in}, \\
&\frac{{d} \hat{\sigma}_{-}}{{d} t} =-\left[{i}\left(\omega_{X^{-}}-\omega\right)+\frac{\gamma}{2}\right] \hat{\sigma}_{-}-g \hat{\sigma}_{z} \hat{a},
\end{aligned}
\end{equation}
where $\omega_{c}$, $\omega_{X^{-}}$, and $\omega$ are the frequencies of the  cavity mode, the ${X^{-}}$  dipole operator, and the incident photon, respectively. $\kappa/2$, $\kappa_{s}/2$, and $\gamma/2$ represent the decay rates of the cavity mode, the side leakage, and  the ${X^{-}}$, respectively.
%
%
Under a weak excitation condition, together  Equation (\ref{eq1}) with the input-output relation  $\hat{a}_{out} =\hat{a}_{in}+\sqrt{\kappa} \; \hat{a}$,  the reflection coefficient for the QD-cavity can be described by\upcite{QD-basic}
\begin{eqnarray}              \label{eq2}
\begin{split}
r_{h}(\omega)=1-\frac{\kappa\left[{i}\left(\omega_{X^{-}}-\omega\right)+\frac{
\gamma}{2}\right]}{\left[{i}\left(\omega_{X^{-}}-\omega\right)+\frac{\gamma}{2}\right]\left[
{i}\left(\omega_{c}-\omega\right)+\frac{\kappa}{2}+\frac{\kappa_{s}}{2}\right]+g^{2}}.
\end{split}
\end{eqnarray}
When QD is uncoupled with the cavity,  i.e., cold cavity with $g=0$, the reflection coefficient becomes
\begin{eqnarray}              \label{eq3}
\begin{split}
r_{0}(\omega)=\frac{{i}\left(\omega_{c}-\omega\right)-\frac{\kappa}{2}+\frac{\kappa_{s}}{2}}{{i}\left(\omega_{c}-\omega\right)+\frac{\kappa}{2}+\frac{\kappa_{s}}{2}}.
\end{split}
\end{eqnarray}
Based on the spin-dependent rules, one can see that the block shown in Figure \ref{singleQD}(c) can be applied in heralded unity-fidelity static, fly, or hybrid quantum computing as it completes the following transformations
\begin{eqnarray}              \label{eq5}
\begin{split}
&&|R\rangle|\pm\rangle \xrightarrow{\text{block$_{1,2,3}$}}\frac{p}{2}(r_h-r_0)|R\rangle|\mp\rangle-(\frac{p}{2}(r_0+r_h)+\sqrt{1-p^2})|L\rangle|\pm\rangle,\\
&&|L\rangle|\pm\rangle \xrightarrow{\text{block$_{1,2,3}$}}\frac{p}{2}(r_h-r_0)|L\rangle|\mp\rangle-(\frac{p}{2}(r_0+r_h)+\sqrt{1-p^2})|R\rangle|\pm\rangle.\\
\end{split}
\end{eqnarray}
Here  $|\pm\rangle\equiv \frac{1}{\sqrt{2}}(|\uparrow\rangle\pm|\downarrow\rangle)$, $p^2$ is the rate of the interaction between QD and incident photon.

\subsection{Complete hyper-Bell states analysis} \label{Sec2.2}

\begin{figure} [tpb]
\begin{center}
\includegraphics[width=11.5 cm,angle=0]{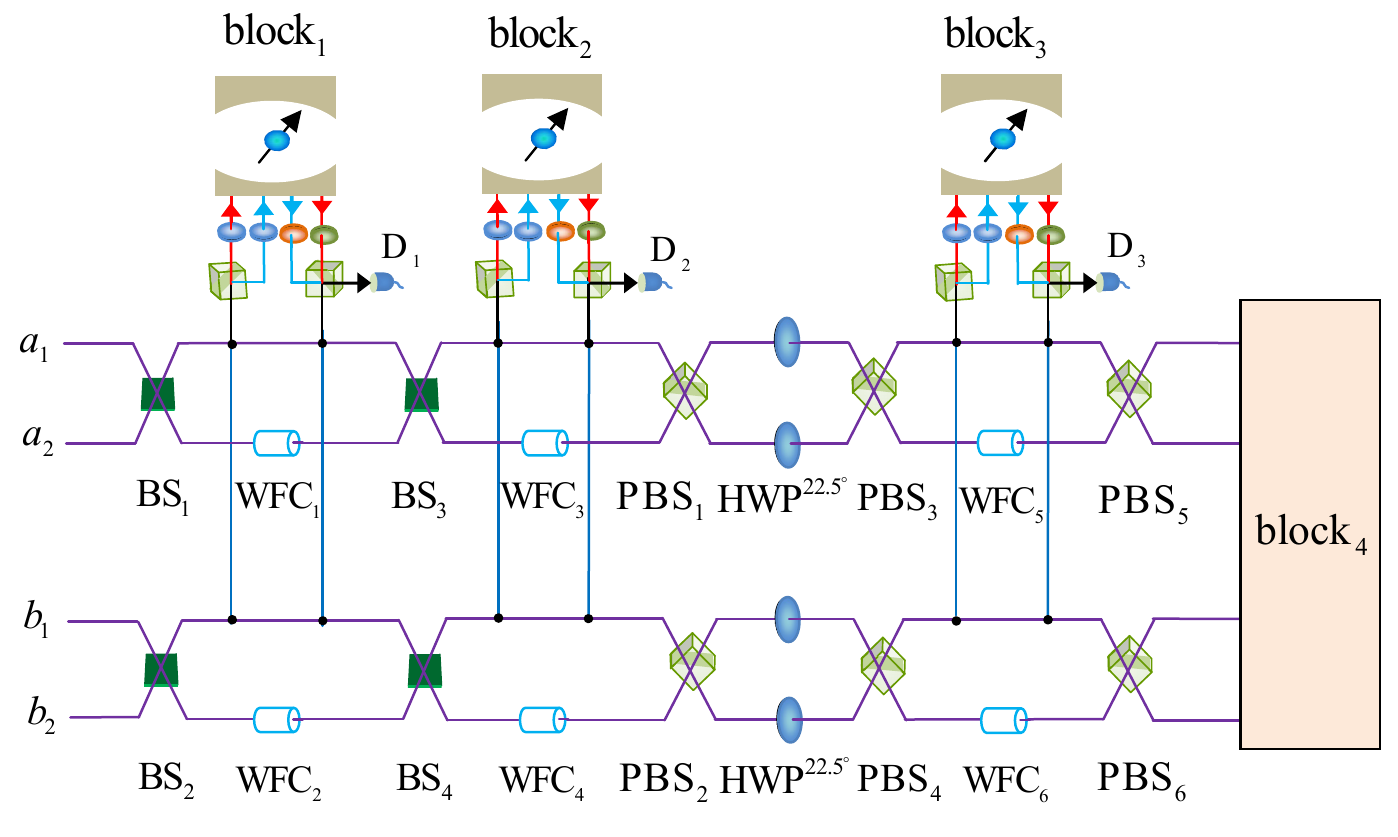}\\\includegraphics[width=10.5 cm,angle=0]{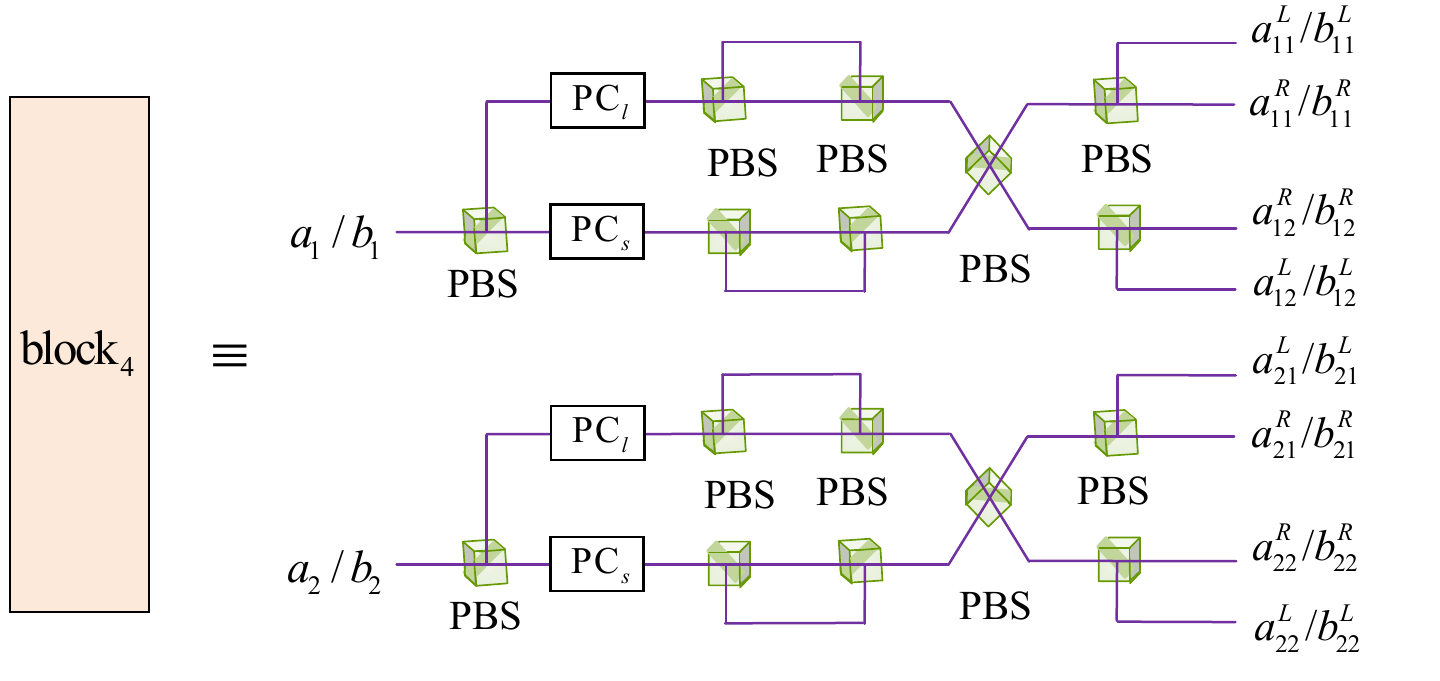}
\caption{Schematic diagram of a complete two-photon three-DOF  hyper-Bell states analysis. $|\text{PC}\rangle_{l }$ ($|\text{PC}\rangle_{s}$) is a Pockel cell which completes a bit-flip operation when the $l$ ($s$) component presents.
50:50 beam splitter, BS, acts exactly as a Hadamard operation for the spatial modes.
Each wave-form corrector, WFC, maps
$|R\rangle\rightarrow \frac{p}{2}(r_h-r_0)|R\rangle$ and
$|L\rangle\rightarrow \frac{p}{2}(r_h-r_0)|L\rangle$.\label{local}}
\end{center}
\end{figure}

Based on above unity-fidelity block depicted by Figure \ref{singleQD}(c), we design a scheme to completely distinguish  two-photon hyper-Bell states in polarization, spatial, and time-bin DOFs, see Figure \ref{local}. Such 64 hyper-Bell states can be written as
\begin{eqnarray}              \label{eq6}
\begin{split}
|\Upsilon\rangle_{AB} = |\Theta_S\rangle_{AB} \otimes |\Gamma_P\rangle_{AB} \otimes |\Xi_T\rangle _{AB}.
\end{split}
\end{eqnarray}
Here $|\Theta _S\rangle _{AB}$ is one of the following four Bell states in the spatial mode:
\begin{eqnarray}              \label{eq7}
\begin{split}
|\phi_S^\pm\rangle_{AB} = \frac{1}{\sqrt{2}}(|a_1 b_1\rangle \pm |a_2 b_2\rangle),\;\;
|\psi_S^\pm\rangle_{AB} = \frac{1}{\sqrt{2}}(|a_1 b_2\rangle \pm |a_2 b_1\rangle).
\end{split}
\end{eqnarray}
$|\Gamma _P\rangle _{AB}$ is one of the  following four Bell states in the  polarization mode:
\begin{eqnarray}              \label{eq8}
\begin{split}
|\phi_P^\pm \rangle _{AB} = \frac{1}{\sqrt{2}}(|RR\rangle \pm |LL\rangle ),\;\;
|\psi_P^\pm \rangle _{AB} = \frac{1}{\sqrt{2}}(|RL\rangle \pm |LR\rangle ).
\end{split}
\end{eqnarray}
$|\Xi _T\rangle _{AB}$ is one of the following four  Bell states in the time-bin mode:
\begin{eqnarray}              \label{eq9}
\begin{split}
|\phi_T^\pm \rangle _{AB} = \frac{1}{\sqrt{2}}{(| ll \rangle \pm |ss \rangle )},\;\;
|\psi_T^\pm \rangle _{AB} = \frac{1}{\sqrt{2}}{(| sl \rangle \pm |ls \rangle )}.
\end{split}
\end{eqnarray}
The subscripts $A$ and $B$ represent the photon $A$ and $B$, respectively. $S$, $P$, and $T$ represent the spatial, the polarization, and the time-bin modes, respectively.
$a_1$ ($b_1$) and $a_2$ ($b_2$) are the two spatial modes of photon $A$ ($B$).
$R$ and $L$ stand for the right-circularly and the left-circularly polarized photons, respectively.
$l$ and $s$ stand for the long time bin and the short time bin, respectively.

Let us follow the process of our proposal, step by step. The states of QD$_1$, QD$_2$, and QD$_3$ contained in block$_1$, block$_2$, and block$_3$ are initially prepared in the states
\begin{eqnarray}              \label{eq10}
\left|\Pi_{\text{block}_1} \right\rangle = \left|+\right\rangle_1,\;
\left|\Pi_{\text{block}_2} \right\rangle = \left|+\right\rangle_2,\;
\left|\Pi_{\text{block}_3} \right\rangle = \left|+\right\rangle_3.
\end{eqnarray}
First, as shown in Figure \ref{local},  photons $A$ and $B$ are injected into the setup in succession, after photon $A$ ($B$) interacts with BS$_1$, block$_1$, WFC$_1$, and BS$_3$ (BS$_2$, block$_1$, WFC$_2$, and BS$_4$) in succession. When the single-photon detector $D_1$ of block$_1$ is not clicked, the states of the composite system will evolve as
\begin{eqnarray}              \label{eq11}
\begin{split}
|\phi_S^+\rangle_{AB} |\Gamma_P\rangle_{AB} |\Xi_T\rangle_{AB} |+\rangle_1 |+\rangle_2 |+\rangle_3
\xrightarrow[\text{BS}_2,\text{block}_1,\text{WFC}_2,\text{BS}_4]{\text{BS}_1,\text{block}_1,\text{WFC}_1,\text{BS}_3}
({p \over 2}({r_h} - {r_0}))^2|\phi_S^+\rangle_{AB} |\Gamma_P\rangle_{AB} |\Xi_T\rangle_{AB} |+\rangle_1|+\rangle_2 |+\rangle_3,
\end{split}
\end{eqnarray}
\begin{eqnarray}              \label{eq12}
\begin{split}
|\psi_S^+\rangle_{AB} |\Gamma_P\rangle_{AB} |\Xi_T\rangle_{AB} |+\rangle_1 |+\rangle_2 |+\rangle_3
\xrightarrow[\text{BS}_2,\text{block}_1,\text{WFC}_2,\text{BS}_4]{\text{BS}_1,\text{block}_1,\text{WFC}_1,\text{BS}_3}
({p \over 2}({r_h} - {r_0}))^2|\psi_S^+\rangle_{AB} |\Gamma_P\rangle_{AB} |\Xi_T\rangle_{AB} |+\rangle_1 |+\rangle_2 |+\rangle_3,
\end{split}
\end{eqnarray}
\begin{eqnarray}              \label{eq13}
\begin{split}
|\phi_S^-\rangle_{AB} |\Gamma_P\rangle_{AB} |\Xi_T\rangle_{AB} |+\rangle_1 |+\rangle_2 |+\rangle_3
\xrightarrow[\text{BS}_2,\text{block}_1,\text{WFC}_2,\text{BS}_4]{\text{BS}_1,\text{block}_1,\text{WFC}_1,\text{BS}_3}
({p \over 2}({r_h} - {r_0}))^2|\phi_S^-\rangle_{AB} |\Gamma_P\rangle_{AB} |\Xi_T\rangle_{AB} |-\rangle_1 |+\rangle_2 |+\rangle_3,
\end{split}
\end{eqnarray}
\begin{eqnarray}              \label{eq14}
\begin{split}
|\psi_S^-\rangle_{AB} |\Gamma_P\rangle_{AB} |\Xi_T\rangle_{AB} |+\rangle_1 |+\rangle_2 |+\rangle_3
\xrightarrow[\text{BS}_2,\text{block}_1,\text{WFC}_2,\text{BS}_4]{\text{BS}_1,\text{block}_1,\text{WFC}_1,\text{BS}_3}
({p \over 2}({r_h} - {r_0}))^2|\psi_S^-\rangle_{AB} |\Gamma_P\rangle_{AB} |\Xi_T\rangle_{AB} |-\rangle_1 |+\rangle_2 |+\rangle_3.
\end{split}
\end{eqnarray}
Here each BS represents an non-polarization 50:50 beam splitter, and it acts exactly as a Hadamard operation for the spatial modes, i.e.,
\begin{eqnarray}              \label{eq15}
\begin{split}
&\left|a_1\right\rangle \xrightarrow{\text{BS}} \frac{1}{\sqrt{2}}(|a_1\rangle+|a_2\rangle),\;\;\;
\left|a_2\right\rangle \xrightarrow{\text{BS}} \frac{1}{\sqrt{2}}(|a_1\rangle-|a_2\rangle),\\
&\left|b_1\right\rangle \xrightarrow{\text{BS}} \frac{1}{\sqrt{2}}(|b_1\rangle+|b_2\rangle),\;\;\;\;
\left|b_2\right\rangle \xrightarrow{\text{BS}} \frac{1}{\sqrt{2}}(|b_1\rangle-|b_2\rangle).
\end{split}
\end{eqnarray}
Based on  Equations (\ref{eq11}-\ref{eq14}), one can see that the spins of QD$_1$ discriminate between $\{|\phi_S^{+}\rangle |\Gamma_P\rangle |\Xi_T\rangle,$ $|\psi_S^+\rangle |\Gamma_P\rangle |\Xi_T\rangle\}$ corresponding to $|+\rangle_1$ and
$\{|\phi_S^{-}\rangle |\Gamma_P\rangle |\Xi_T\rangle$, $|\psi_S^-\rangle |\Gamma_P\rangle |\Xi_T\rangle\}$ corresponding to $|-\rangle_1$.

Second, the photons pass through block$_2$, WFC$_3$, PBS$_1$ (block$_2$, WFC$_4$, PBS$_2$) in succession. When the single-photon detector $D_2$ of block$_2$ is not clicked, the transformations of the states described by Equations (\ref{eq11}-\ref{eq14}) can be written as
\begin{eqnarray}              \label{eq16}
\begin{split}
({p \over 2}({r_h} - {r_0}))^2|\phi_S^{+}\rangle_{AB} |\phi_P^\pm\rangle_{AB} |\Xi_T\rangle_{AB} |+\rangle_1 |+\rangle_2 |+\rangle_3
\xrightarrow[\text{block}_2,\text{WFC}_4,\text{PBS}_2]{\text{block}_2,\text{WFC}_3,\text{PBS}_1}
&({p \over 2}({r_h} - {r_0}))^4|\phi_S^{+}\rangle_{AB} |\phi_P^\pm\rangle_{AB}|\Xi_T\rangle_{AB}
\\&\otimes|+\rangle_1 |+\rangle_2 |+\rangle_3,
\end{split}
\end{eqnarray}
\begin{eqnarray}              \label{eq17}
\begin{split}
({p \over 2}({r_h} - {r_0}))^2|\psi_S^{+}\rangle_{AB} |\phi_P^\pm\rangle_{AB} |\Xi_T\rangle_{AB} |+\rangle_1 |+\rangle_2 |+\rangle_3
\xrightarrow[\text{block}_2,\text{WFC}_4,\text{PBS}_2]{\text{block}_2,\text{WFC}_3,\text{PBS}_1}
&({p \over 2}({r_h} - {r_0}))^4|\psi_S^{+}\rangle_{AB} |\phi_P^\pm\rangle_{AB}|\Xi_T\rangle_{AB}\\&\otimes
|+\rangle_1 |-\rangle_2 |+\rangle_3,
\end{split}
\end{eqnarray}
\begin{eqnarray}              \label{eq18}
\begin{split}
({p \over 2}({r_h} - {r_0}))^2|\phi_S^{+}\rangle_{AB} |\psi_P^\pm\rangle_{AB} |\Xi_T\rangle_{AB} |+\rangle_1 |+\rangle_2 |+\rangle_3
\xrightarrow[\text{block}_2,\text{WFC}_4,\text{PBS}_2]{\text{block}_2,\text{WFC}_3,\text{PBS}_1}
&({p \over 2}({r_h} - {r_0}))^4|\psi_S^{+}\rangle_{AB} |\psi_P^\pm\rangle_{AB} |\Xi_T\rangle_{AB}\\&\otimes
|+\rangle_1 |+\rangle_2 |+\rangle_3,
\end{split}
\end{eqnarray}
\begin{eqnarray}              \label{eq19}
\begin{split}
({p \over 2}({r_h} - {r_0}))^2|\psi_S^{+}\rangle_{AB} |\psi_P^\pm\rangle_{AB} |\Xi_T\rangle_{AB} |+\rangle_1 |+\rangle_2 |+\rangle_3
\xrightarrow[\text{block}_2,\text{WFC}_4,\text{PBS}_2]{\text{block}_2,\text{WFC}_3,\text{PBS}_1}
&({p \over 2}({r_h} - {r_0}))^4|\phi_S^{+}\rangle_{AB} |\psi_P^\pm\rangle_{AB}|\Xi_T\rangle_{AB}\\&\otimes
|+\rangle_1 |-\rangle_2 |+\rangle_3,
\end{split}
\end{eqnarray}
\begin{eqnarray}              \label{eq20}
\begin{split}
({p \over 2}({r_h} - {r_0}))^2|\phi_S^{-}\rangle_{AB} |\phi_P^\pm\rangle_{AB} |\Xi_T\rangle_{AB} |-\rangle_1 |+\rangle_2 |+\rangle_3
\xrightarrow[\text{block}_2,\text{WFC}_4,\text{PBS}_2]{\text{block}_2,\text{WFC}_3,\text{PBS}_1}
&({p \over 2}({r_h} - {r_0}))^4|\phi_S^{-}\rangle_{AB} |\phi_P^\mp\rangle_{AB}|\Xi_T\rangle_{AB}\\&\otimes
|-\rangle_1 |+\rangle_2 |+\rangle_3,
\end{split}
\end{eqnarray}
\begin{eqnarray}              \label{eq21}
\begin{split}
({p \over 2}({r_h} - {r_0}))^2|\psi_S^{-}\rangle_{AB} |\phi_P^\pm\rangle_{AB} |\Xi_T\rangle_{AB} |-\rangle_1 |+\rangle_2 |+\rangle_3
\xrightarrow[\text{block}_2,\text{WFC}_4,\text{PBS}_2]{\text{block}_2,\text{WFC}_3,\text{PBS}_1}
&({p \over 2}({r_h} - {r_0}))^4|\psi_S^{-}\rangle_{AB} |\phi_P^\mp\rangle_{AB}|\Xi_T\rangle_{AB}
\\&\otimes|-\rangle_1 |-\rangle_2 |+\rangle_3,
\end{split}
\end{eqnarray}%
\begin{eqnarray}              \label{eq22}
\begin{split}
({p \over 2}({r_h} - {r_0}))^2|\phi_S^{-}\rangle_{AB} |\psi_P^\pm\rangle_{AB} |\Xi_T\rangle_{AB} |-\rangle_1 |+\rangle_2 |+\rangle_3
\xrightarrow[\text{block}_2,\text{WFC}_4,\text{PBS}_2]{\text{block}_2,\text{WFC}_3,\text{PBS}_1}
&({p \over 2}({r_h} - {r_0}))^4|\psi_S^{-}\rangle_{AB} |\psi_P^\mp\rangle_{AB}|\Xi_T\rangle_{AB}
\\&\otimes|-\rangle_1 |+\rangle_2 |+\rangle_3,
\end{split}
\end{eqnarray}
\begin{eqnarray}              \label{eq23}
\begin{split}
({p \over 2}({r_h} - {r_0}))^2|\psi_S^{-}\rangle_{AB} |\psi_P^\pm\rangle_{AB} |\Xi_T\rangle_{AB} |-\rangle_1 |+\rangle_2 |+\rangle_3
\xrightarrow[\text{block}_2,\text{WFC}_4,\text{PBS}_2]{\text{block}_2,\text{WFC}_3,\text{PBS}_1}
&({p \over 2}({r_h} - {r_0}))^4|\phi_S^{-}\rangle_{AB} |\psi_P^\mp\rangle_{AB}|\Xi_T\rangle_{AB}\\&\otimes
|-\rangle_1 |-\rangle_2 |+\rangle_3.
\end{split}
\end{eqnarray}
Then, the 64 hyper-Bell states can be divided into the 4 distinct groups according to the spins of QD$_1$ and QD$_2$. That is,
\{$|\phi_S^{+}\rangle_{AB} |\phi_P^\pm\rangle_{AB}|\Xi_T\rangle_{AB}, |\psi_S^{+}\rangle_{AB} |\psi_P^\pm\rangle_{AB} |\Xi_T\rangle_{AB}\}$ corresponds to $|+\rangle_1 |+\rangle_2$,
$\{|\psi_S^{+}\rangle_{AB} |\phi_P^\pm\rangle_{AB}|\Xi_T\rangle_{AB}, |\phi_S^{+}\rangle_{AB} |\psi_P^\pm\rangle_{AB}|\Xi_T\rangle_{AB}\}$ corresponds to $|+\rangle_1 |-\rangle_2$,
$\{|\phi_S^{-}\rangle_{AB} |\phi_P^\mp\rangle_{AB}|\Xi_T\rangle_{AB}, |\psi_S^{-}\rangle_{AB} |\psi_P^\mp\rangle_{AB}|\Xi_T\rangle_{AB}\}$ corresponds to $|-\rangle_1 |+\rangle_2$,
and
$\{|\psi_S^{-}\rangle_{AB} |\phi_P^\mp\rangle_{AB}|\Xi_T\rangle_{AB}, |\phi_S^{-}\rangle_{AB} |\psi_P^\mp\rangle_{AB}|\Xi_T\rangle_{AB}\}$ corresponds to $|-\rangle_1 |-\rangle_2$.

Third, the photons pass through HWP$^{22.5^\circ}$, PBS$_3$, block$_3$, WFC$_5$, and PBS$_5$ (HWP$^{22.5^\circ}$, PBS$_4$, block$_3$, WFC$_6$, and PBS$_6$) in succession.
When the single-photon detector $D_3$ of block$_3$ is not clicked,  the states of the composite system becomes
\begin{eqnarray}              \label{eq24}
\begin{split}
({p \over 2}({r_h} - {r_0}))^4|\phi_S^+\rangle_{AB} |\phi_P^+\rangle_{AB} |\Xi_T\rangle_{AB} |+\rangle_1 |+\rangle_2 |+\rangle_3
\xrightarrow[\text{HWP}_1^{22.5^\circ},\text{PBS}_4,\text{block}_3,\text{WFC}_6,\text{PBS}_6]
{\text{HWP}_1^{22.5^\circ},\text{PBS}_3,\text{block}_3,\text{WFC}_5,\text{PBS}_5}
&({p \over 2}({r_h} - {r_0}))^6|\phi_S^+\rangle_{AB} |\phi_P^+\rangle_{AB} \\
&\otimes
|\Xi_T\rangle_{AB}|+\rangle_1 |+\rangle_2 |+\rangle_3,
\end{split}
\end{eqnarray}
\begin{eqnarray}              \label{eq25}
\begin{split}
({p \over 2}({r_h} - {r_0}))^4|\phi_S^+\rangle_{AB} |\phi_P^-\rangle_{AB} |\Xi_T\rangle_{AB} |+\rangle_1 |+\rangle_2 |+\rangle_3
\xrightarrow[\text{HWP}_1^{22.5^\circ},\text{PBS}_4,\text{block}_3,\text{WFC}_6,\text{PBS}_6]
{\text{HWP}_1^{22.5^\circ},\text{PBS}_3,\text{block}_3,\text{WFC}_5,\text{PBS}_5}
&({p \over 2}({r_h} - {r_0}))^6|\phi_S^+\rangle_{AB} |\psi_P^+\rangle_{AB}\\
&\otimes
 |\Xi_T\rangle_{AB}|+\rangle_1 |+\rangle_2 |-\rangle_3,
\end{split}
\end{eqnarray}
\begin{eqnarray}              \label{eq26}
\begin{split}
({p \over 2}({r_h} - {r_0}))^4|\psi_S^+\rangle_{AB} |\psi_P^+\rangle_{AB} |\Xi_T\rangle_{AB} |+\rangle_1 |+\rangle_2 |+\rangle_3
\xrightarrow[\text{HWP}_1^{22.5^\circ},\text{PBS}_4,\text{block}_3,\text{WFC}_6,\text{PBS}_6]
{\text{HWP}_1^{22.5^\circ},\text{PBS}_3,\text{block}_3,\text{WFC}_5,\text{PBS}_5}
&({p \over 2}({r_h} - {r_0}))^6|\psi_S^+\rangle_{AB} |\phi_P^-\rangle_{AB}\\
&\otimes
|\Xi_T\rangle_{AB}|+\rangle_1 |+\rangle_2 |-\rangle_3,
\end{split}
\end{eqnarray}
\begin{eqnarray}              \label{eq27}
\begin{split}
({p \over 2}({r_h} - {r_0}))^4|\psi_S^+\rangle_{AB} |\psi_P^-\rangle_{AB} |\Xi_T\rangle_{AB} |+\rangle_1 |+\rangle_2 |+\rangle_3
\xrightarrow[\text{HWP}_1^{22.5^\circ},\text{PBS}_4,\text{block}_3,\text{WFC}_6,\text{PBS}_6]
{\text{HWP}_1^{22.5^\circ},\text{PBS}_3,\text{block}_3,\text{WFC}_5,\text{PBS}_5}
&({p \over 2}({r_h} - {r_0}))^6|\psi_S^+\rangle_{AB} |\psi_P^-\rangle_{AB}\\
&\otimes
|\Xi_T\rangle_{AB}|+\rangle_1 |+\rangle_2 |+\rangle_3,
\end{split}
\end{eqnarray}
\begin{eqnarray}              \label{eq28}
\begin{split}
({p \over 2}({r_h} - {r_0}))^4|\phi_S^+\rangle_{AB} |\psi_P^+\rangle_{AB} |\Xi_T\rangle_{AB} |+\rangle_1 |-\rangle_2 |+\rangle_3
\xrightarrow[\text{HWP}_1^{22.5^\circ},\text{PBS}_4,\text{block}_3,\text{WFC}_6,\text{PBS}_6]
{\text{HWP}_1^{22.5^\circ},\text{PBS}_3,\text{block}_3,\text{WFC}_5,\text{PBS}_5}
&({p \over 2}({r_h} - {r_0}))^6|\phi_S^+\rangle_{AB} |\phi_P^-\rangle_{AB}\\
&\otimes
|\Xi_T\rangle_{AB}|+\rangle_1 |-\rangle_2 |+\rangle_3,
\end{split}
\end{eqnarray}
\begin{eqnarray}              \label{eq29}
\begin{split}
({p \over 2}({r_h} - {r_0}))^4|\phi_S^+\rangle_{AB} |\psi_P^-\rangle_{AB} |\Xi_T\rangle_{AB} |+\rangle_1 |-\rangle_2 |+\rangle_3
\xrightarrow[\text{HWP}_1^{22.5^\circ},\text{PBS}_4,\text{block}_3,\text{WFC}_6,\text{PBS}_6]
{\text{HWP}_1^{22.5^\circ},\text{PBS}_3,\text{block}_3,\text{WFC}_5,\text{PBS}_5}
&({p \over 2}({r_h} - {r_0}))^6|\phi_S^+\rangle_{AB} |\psi_P^-\rangle_{AB}\\
&\otimes
|\Xi_T\rangle_{AB}|+\rangle_1 |-\rangle_2 |-\rangle_3,
\end{split}
\end{eqnarray}
\begin{eqnarray}              \label{eq30}
\begin{split}
({p \over 2}({r_h} - {r_0}))^4|\psi_S^+\rangle_{AB} |\phi_P^+\rangle_{AB} |\Xi_T\rangle_{AB} |+\rangle_1 |-\rangle_2 |+\rangle_3
\xrightarrow[\text{HWP}_1^{22.5^\circ},\text{PBS}_4,\text{block}_3,\text{WFC}_6,\text{PBS}_6]
{\text{HWP}_1^{22.5^\circ},\text{PBS}_3,\text{block}_3,\text{WFC}_5,\text{PBS}_5}
&({p \over 2}({r_h} - {r_0}))^6|\psi_S^+\rangle_{AB} |\phi_P^+\rangle_{AB}\\
&\otimes
|\Xi_T\rangle_{AB}|+\rangle_1 |-\rangle_2 |-\rangle_3,
\end{split}
\end{eqnarray}
\begin{eqnarray}              \label{eq31}
\begin{split}
({p \over 2}({r_h} - {r_0}))^4|\psi_S^+\rangle_{AB} |\phi_P^-\rangle_{AB} |\Xi_T\rangle_{AB} |+\rangle_1 |-\rangle_2 |+\rangle_3
\xrightarrow[\text{HWP}_1^{22.5^\circ},\text{PBS}_4,\text{block}_3,\text{WFC}_6,\text{PBS}_6]
{\text{HWP}_1^{22.5^\circ},\text{PBS}_3,\text{block}_3,\text{WFC}_5,\text{PBS}_5}
&({p \over 2}({r_h} - {r_0}))^6|\psi_S^+\rangle_{AB} |\psi_P^+\rangle_{AB}\\
&\otimes
|\Xi_T\rangle_{AB}|+\rangle_1 |-\rangle_2 |+\rangle_3,
\end{split}
\end{eqnarray}
\begin{eqnarray}              \label{eq32}
\begin{split}
({p \over 2}({r_h} - {r_0}))^4|\phi_S^-\rangle_{AB} |\phi_P^+\rangle_{AB} |\Xi_T\rangle_{AB} |-\rangle_1 |+\rangle_2 |+\rangle_3
\xrightarrow[\text{HWP}_1^{22.5^\circ},\text{PBS}_4,\text{block}_3,\text{WFC}_6,\text{PBS}_6]
{\text{HWP}_1^{22.5^\circ},\text{PBS}_3,\text{block}_3,\text{WFC}_5,\text{PBS}_5}
&({p \over 2}({r_h} - {r_0}))^6|\phi_S^-\rangle_{AB} |\phi_P^+\rangle_{AB}\\
&\otimes
 |\Xi_T\rangle_{AB}|-\rangle_1 |+\rangle_2 |+\rangle_3,
\end{split}
\end{eqnarray}
\begin{eqnarray}              \label{eq33}
\begin{split}
({p \over 2}({r_h} - {r_0}))^4|\phi_S^-\rangle_{AB} |\phi_P^-\rangle_{AB} |\Xi_T\rangle_{AB} |-\rangle_1 |+\rangle_2 |+\rangle_3
\xrightarrow[\text{HWP}_1^{22.5^\circ},\text{PBS}_4,\text{block}_3,\text{WFC}_6,\text{PBS}_6]
{\text{HWP}_1^{22.5^\circ},\text{PBS}_3,\text{block}_3,\text{WFC}_5,\text{PBS}_5}
&({p \over 2}({r_h} - {r_0}))^6|\phi_S^-\rangle_{AB} |\psi_P^+\rangle_{AB}\\
&\otimes
|\Xi_T\rangle_{AB} |-\rangle_1 |+\rangle_2 |-\rangle_3,
\end{split}
\end{eqnarray}

\begin{eqnarray}              \label{eq34}
\begin{split}
({p \over 2}({r_h} - {r_0}))^4|\psi_S^-\rangle_{AB} |\psi_P^+\rangle_{AB} |\Xi_T\rangle_{AB} |-\rangle_1 |+\rangle_2 |+\rangle_3
\xrightarrow[\text{HWP}_1^{22.5^\circ},\text{PBS}_4,\text{block}_3,\text{WFC}_6,\text{PBS}_6]
{\text{HWP}_1^{22.5^\circ},\text{PBS}_3,\text{block}_3,\text{WFC}_5,\text{PBS}_5}
&({p \over 2}({r_h} - {r_0}))^6|\psi_S^-\rangle_{AB} |\phi_P^-\rangle_{AB}\\
&\otimes
|\Xi_T\rangle_{AB}|-\rangle_1 |+\rangle_2 |-\rangle_3,
\end{split}
\end{eqnarray}
\begin{eqnarray}              \label{eq35}
\begin{split}
({p \over 2}({r_h} - {r_0}))^4|\psi_S^-\rangle_{AB} |\psi_P^-\rangle_{AB} |\Xi_T\rangle_{AB} |-\rangle_1 |+\rangle_2 |+\rangle_3
\xrightarrow[\text{HWP}_1^{22.5^\circ},\text{PBS}_4,\text{block}_3,\text{WFC}_6,\text{PBS}_6]
{\text{HWP}_1^{22.5^\circ},\text{PBS}_3,\text{block}_3,\text{WFC}_5,\text{PBS}_5}
&({p \over 2}({r_h} - {r_0}))^6|\psi_S^-\rangle_{AB} |\psi_P^-\rangle_{AB}\\
&\otimes|\Xi_T\rangle_{AB}|-\rangle_1 |+\rangle_2 |+\rangle_3,
\end{split}
\end{eqnarray}

\begin{eqnarray}              \label{eq36}
\begin{split}
({p \over 2}({r_h} - {r_0}))^4|\phi_S^-\rangle_{AB} |\psi_P^+\rangle_{AB} |\Xi_T\rangle_{AB} |-\rangle_1 |-\rangle_2 |+\rangle_3
\xrightarrow[\text{HWP}_1^{22.5^\circ},\text{PBS}_4,\text{block}_3,\text{WFC}_6,\text{PBS}_6]
{\text{HWP}_1^{22.5^\circ},\text{PBS}_3,\text{block}_3,\text{WFC}_5,\text{PBS}_5}
&({p \over 2}({r_h} - {r_0}))^6|\phi_S^-\rangle_{AB}|\phi_P^-\rangle_{AB}\\
&\otimes
|\Xi_T\rangle_{AB}|-\rangle_1 |-\rangle_2 |+\rangle_3,
\end{split}
\end{eqnarray}
\begin{eqnarray}              \label{eq37}
\begin{split}
({p \over 2}({r_h} - {r_0}))^4|\phi_S^-\rangle_{AB} |\psi_P^-\rangle_{AB} |\Xi_T\rangle_{AB} |-\rangle_1 |-\rangle_2 |+\rangle_3
\xrightarrow[\text{HWP}_1^{22.5^\circ},\text{PBS}_4,\text{block}_3,\text{WFC}_6,\text{PBS}_6]
{\text{HWP}_1^{22.5^\circ},\text{PBS}_3,\text{block}_3,\text{WFC}_5,\text{PBS}_5}
&({p \over 2}({r_h} - {r_0}))^6|\phi_S^-\rangle_{AB} |\psi_P^-\rangle_{AB}\\
&\otimes
|\Xi_T\rangle_{AB}|-\rangle_1 |-\rangle_2 |-\rangle_3,
\end{split}
\end{eqnarray}
\begin{eqnarray}              \label{eq38}
\begin{split}
({p \over 2}({r_h} - {r_0}))^4|\psi_S^-\rangle_{AB} |\phi_P^+\rangle_{AB} |\Xi_T\rangle_{AB} |-\rangle_1 |-\rangle_2 |+\rangle_3
\xrightarrow[\text{HWP}_1^{22.5^\circ},\text{PBS}_4,\text{block}_3,\text{WFC}_6,\text{PBS}_6]
{\text{HWP}_1^{22.5^\circ},\text{PBS}_3,\text{block}_3,\text{WFC}_5,\text{PBS}_5}
&({p \over 2}({r_h} - {r_0}))^6|\psi_S^-\rangle_{AB} |\phi_P^+\rangle_{AB}\\
&\otimes
|\Xi_T\rangle_{AB}|-\rangle_1 |-\rangle_2 |-\rangle_3,
\end{split}
\end{eqnarray}
\begin{eqnarray}              \label{eq39}
\begin{split}
({p \over 2}({r_h} - {r_0}))^4|\psi_S^-\rangle_{AB} |\phi_P^-\rangle_{AB} |\Xi_T\rangle_{AB} |-\rangle_1 |-\rangle_2 |+\rangle_3
\xrightarrow[\text{HWP}_1^{22.5^\circ},\text{PBS}_4,\text{block}_3,\text{WFC}_6,\text{PBS}_6]
{\text{HWP}_1^{22.5^\circ},\text{PBS}_3,\text{block}_3,\text{WFC}_5,\text{PBS}_5}
&({p \over 2}({r_h} - {r_0}))^6|\psi_S^-\rangle_{AB} |\psi_P^+\rangle_{AB}\\
&\otimes
|\Xi_T\rangle_{AB}|-\rangle_1 |-\rangle_2 |+\rangle_3.
\end{split}
\end{eqnarray}
Based on Equations (\ref{eq6}-\ref{eq14}), Equations (\ref{eq16}-\ref{eq23}), and Equations (\ref{eq24}-\ref{eq39}), one can see that the spins of QD$_1$, QD$_2$, and QD$_3$ perform the polarization and spatial hyper-Bell states analysis. That is, as shown in Table \ref{table1}, 8 distinguishable groups
$\{|\phi_S^+\rangle|\phi_P^+\rangle|\Xi_T\rangle_{AB}$, $|\phi_S^+\rangle|\psi_P^-\rangle|\Xi_T\rangle_{AB}\}$,
$\{|\phi_S^+\rangle|\phi_P^-\rangle|\Xi_T\rangle_{AB}$, $|\phi_S^+\rangle|\psi_P^+\rangle|\Xi_T\rangle_{AB}\}$,
$\{|\psi_S^+\rangle|\phi_P^-\rangle|\Xi_T\rangle_{AB}$, $|\psi_S^+\rangle|\psi_P^+\rangle|\Xi_T\rangle_{AB}\}$,
$\{|\psi_S^+\rangle|\phi_P^+\rangle|\Xi_T\rangle_{AB}$, $|\psi_S^+\rangle|\psi_P^-\rangle|\Xi_T\rangle_{AB}\}$,
$\{|\phi_S^-\rangle|\phi_P^-\rangle|\Xi_T\rangle_{AB}$, $|\phi_S^-\rangle|\psi_P^+\rangle|\Xi_T\rangle_{AB}\}$,
$\{|\phi_S^-\rangle|\phi_P^-\rangle|\Xi_T\rangle_{AB}$, $|\phi_S^-\rangle|\psi_P^-\rangle|\Xi_T\rangle_{AB}\}$,
$\{|\psi_S^-\rangle|\phi_P^+\rangle_|\Xi_T\rangle_{AB}$, $|\psi_S^-\rangle|\psi_P^-\rangle|\Xi_T\rangle_{AB}\}$, and
$\{|\psi_S^-\rangle|\phi_P^-\rangle|\Xi_T\rangle_{AB}$, $|\psi_S^-\rangle|\psi_P^+\rangle|\Xi_T\rangle_{AB}\}$ can be obtained.

Fourth, the photons pass through the block$_4$, which composed of sequences of ``PBSs and PCs'', shown in Figure \ref{local}. Here the block$_4$ induces the following transformations:
\begin{eqnarray}              \label{eq40}
\begin{split}
({p \over 2}({r_h} - {r_0}))^6|\phi_S^\pm\rangle_{AB} |\phi_P^+\rangle_{AB}|\phi_T^+\rangle_{AB}& |\pm\rangle_1|+\rangle_2 |+\rangle_3
\xrightarrow{\text{block}_4}
{1 \over {2\sqrt {2}}}({p \over 2}({r_h} - {r_0}))^6(|a_{11}^Rb_{11}^R\rangle + |a_{12}^R b_{12}^R\rangle + |a_{11}^L b_{11}^L\rangle\\&+|a_{12}^L b_{12}^L\rangle
\pm(|a_{21}^R b_{21}^R\rangle + |a_{22}^R b_{22}^R\rangle + |a_{22}^L b_{22}^L\rangle + |a_{22}^L b_{21}^L\rangle)) |ls,ls\rangle |\pm\rangle_1|+\rangle_2 |+\rangle_3,
\end{split}
\end{eqnarray}
\begin{eqnarray}              \label{eq41}
\begin{split}
({p \over 2}({r_h} - {r_0}))^6|\phi_S^\pm\rangle_{AB}|\phi_P^+\rangle_{AB}|\phi_T^-\rangle_{AB} & |\pm\rangle_1|+\rangle_2 |+\rangle_3
\xrightarrow{\text{block}_4}
{1 \over {2\sqrt {2}}}({p \over 2}({r_h} - {r_0}))^6(|a_{11}^R b_{11}^L\rangle +|a_{12}^R b_{12}^L\rangle + |a_{11}^L b_{11}^R\rangle \\&+ |a_{12}^L b_{12}^R\rangle
\pm(|a_{21}^R b_{21}^L\rangle + |a_{21}^L b_{21}^R\rangle + |a_{22}^R b_{22}^L\rangle + |a_{22}^L b_{21}^R\rangle))|ls,ls\rangle|\pm\rangle_1|+\rangle_2|+\rangle_3,
\end{split}
\end{eqnarray}
\begin{eqnarray}              \label{eq42}
\begin{split}
({p \over 2}({r_h} - {r_0}))^6|\phi_S^\pm\rangle_{AB} |\phi_P^+\rangle_{AB}|\psi_T^+\rangle_{AB} & |\pm\rangle_1|+\rangle_2 |+\rangle_3
\xrightarrow{\text{block}_4}
{1 \over {2\sqrt {2}}}({p \over 2}({r_h} - {r_0}))^6(|a_{11}^R b_{12}^R\rangle+|a_{12}^R b_{11}^R\rangle-|a_{11}^L b_{12}^L\rangle\\& - |a_{12}^L b_{11}^L\rangle
\pm(|a_{21}^R b_{22}^R\rangle + |a_{22}^R b_{21}^R\rangle - |a_{22}^L b_{21}^L\rangle -|a_{22}^L b_{22}^L\rangle))|ls,ls\rangle|\pm\rangle_1|+\rangle_2|+\rangle_3,
\end{split}
\end{eqnarray}

\begin{eqnarray}              \label{eq43}
\begin{split}
({p \over 2}({r_h} - {r_0}))^6|\phi_S^\pm\rangle_{AB} |\phi_P^+\rangle_{AB} |\psi_T^-\rangle_{AB} &|\pm\rangle_1|+\rangle_2 |+\rangle_3
\xrightarrow{\text{block}_4}
{1 \over {2\sqrt {2}}}({p \over 2}({r_h} - {r_0}))^6(|a_{11}^R b_{12}^L\rangle + |a_{12}^R b_{11}^L\rangle - |a_{11}^L b_{12}^R\rangle\\&- |a_{12}^L b_{11}^R\rangle
\pm(|a_{21}^R b_{22}^L\rangle + |a_{22}^R b_{21}^L\rangle - |a_{22}^L b_{21}^R\rangle -|a_{22}^L b_{22}^R\rangle))|ls,ls\rangle |\pm \rangle_1|+\rangle_2|+\rangle_3,
\end{split}
\end{eqnarray}
\begin{eqnarray}              \label{eq44}
\begin{split}
({p \over 2}({r_h} - {r_0}))^6|\psi_S^\pm\rangle_{AB} |\psi_P^-\rangle_{AB} |\phi_T^+\rangle_{AB} & |\pm\rangle_1|+\rangle_2 |+\rangle_3
\xrightarrow{\text{block}_4}
{1 \over {2\sqrt {2}}}({p \over 2}({r_h} - {r_0}))^6(|a_{11}^R b_{22}^L\rangle-|a_{12}^R b_{21}^L\rangle-|a_{12}^L b_{21}^R\rangle\\&+| a_{11}^L b_{22}^R\rangle
\pm(|a_{21}^R b_{12}^L\rangle - |a_{22}^R b_{11}^L\rangle-|a_{22}^L b_{11}^R\rangle+|a_{21}^L b_{12}^R\rangle))|ls,ls\rangle|\pm\rangle_1|+\rangle_2|\rm{+}\rangle_3,
\end{split}
\end{eqnarray}
\begin{eqnarray}              \label{eq45}
\begin{split}
({p \over 2}({r_h} - {r_0}))^6|\psi_S^\pm\rangle_{AB} |\psi_P^-\rangle_{AB} |\phi_T^-\rangle_{AB} & |\pm\rangle_1|+\rangle_2 |+\rangle_3
\xrightarrow{\text{block}_4}
{1 \over {2\sqrt {2}}}({p \over 2}({r_h} - {r_0}))^6(|a_{11}^R b_{22}^R\rangle-|a_{12}^R b_{21}^R\rangle-|a_{12}^L b_{21}^L\rangle\\&+|a_{11}^L b_{22}^L\rangle
\pm(|a_{21}^Rb_{12}^R\rangle -|a_{22}^R b_{11}^R\rangle- |a_{22}^L b_{11}^L\rangle + |a_{21}^Lb_{12}^L\rangle))|ls,ls\rangle|\pm\rangle_1|+\rangle_2|\rm{+}\rangle_3,
\end{split}
\end{eqnarray}

\begin{eqnarray}              \label{eq46}
\begin{split}
({p \over 2}({r_h} - {r_0}))^6|\psi_S^\pm\rangle_{AB} |\psi_P^-\rangle_{AB} |\psi_T^+\rangle_{AB} & |\pm\rangle_1|+\rangle_2 |+\rangle_3
\xrightarrow{\text{block}_4}
{1 \over {2\sqrt {2}}}({p \over 2}({r_h} - {r_0}))^6(|a_{11}^R b_{21}^L\rangle-|a_{12}^Rb_{22}^L\rangle-|a_{12}^Lb_{22}^R\rangle\\&+|a_{11}^L b_{21}^R\rangle
\pm(|a_{21}^R b_{11}^L\rangle - |a_{22}^R b_{12}^L\rangle- |a_{22}^L b_{12}^R\rangle + |a_{21}^L b_{11}^R\rangle))|ls,ls\rangle |\pm\rangle_1|+\rangle_2| \rm{+}\rangle_3,
\end{split}
\end{eqnarray}
\begin{eqnarray}              \label{eq47}
\begin{split}
({p \over 2}({r_h} - {r_0}))^6|\psi_S^\pm\rangle_{AB} |\psi_P^-\rangle_{AB} |\psi_T^-\rangle_{AB} & |\pm\rangle_1|+\rangle_2 |+\rangle_3
\xrightarrow{\text{block}_4}
{1 \over {2\sqrt {2}}}({p \over 2}({r_h} - {r_0}))^6(|a_{11}^R b_{21}^R\rangle-|a_{12}^R b_{22}^R\rangle-|a_{12}^L b_{22}^L\rangle\\&+|a_{11}^L b_{21}^L\rangle
\pm(|a_{21}^R b_{11}^R\rangle-|a_{22}^R b_{12}^R\rangle- |a_{22}^L b_{12}^L\rangle + |a_{21}^L b_{11}^L\rangle))|ls, ls\rangle|\pm\rangle_1|+\rangle_2|\rm{+}\rangle_3,
\end{split}
\end{eqnarray}
\begin{eqnarray}              \label{eq56}
\begin{split}
({p \over 2}({r_h} - {r_0}))^6|\phi_S^\pm\rangle_{AB} |\psi_P^+\rangle_{AB} |\phi_T^+\rangle_{AB} & |\pm\rangle_1|+\rangle_2 |-\rangle_3
\xrightarrow{\text{block}_4}
{1 \over {2\sqrt {2}}}({p \over 2}({r_h} - {r_0}))^6(|a_{11}^Rb_{12}^R\rangle+ |a_{12}^Rb_{11}^R\rangle-|a_{11}^Lb_{12}^L\rangle\\&-|a_{12}^Lb_{11}^L\rangle
\pm(|a_{21}^Rb_{22}^R\rangle + |a_{22}^Rb_{21}^R\rangle- |a_{22}^Lb_{21}^L\rangle -|a_{22}^Lb_{22}^L\rangle))|ls,ls\rangle {|\pm\rangle_1|+\rangle_2|-\rangle_3},
\end{split}
\end{eqnarray}
\begin{eqnarray}              \label{eq57}
\begin{split}
({p \over 2}({r_h} - {r_0}))^6|\phi_S^\pm\rangle_{AB} |\psi_P^+\rangle_{AB} |\phi_T^-\rangle_{AB} & |\pm\rangle_1|+\rangle_2 |-\rangle_3
\xrightarrow{\text{block}_4}
{1 \over {2\sqrt {2}}}({p \over 2}({r_h} - {r_0}))^6(|a_{11}^R b_{12}^L\rangle+ |a_{12}^R b_{11}^L\rangle-|a_{11}^L b_{12}^R\rangle\\&-|a_{12}^L b_{11}^R\rangle
\pm(|a_{21}^R b_{22}^L\rangle+|a_{22}^R b_{21}^L\rangle- |a_{22}^L b_{21}^R\rangle-|a_{22}^L b_{22}^R\rangle))|ls,ls\rangle |\pm\rangle_1|+\rangle_2|-\rangle_3,
\end{split}
\end{eqnarray}
\begin{eqnarray}              \label{eq58}
\begin{split}
({p \over 2}({r_h} - {r_0}))^6|\phi_S^\pm\rangle_{AB} |\psi_P^+\rangle_{AB} |\psi_T^+\rangle_{AB} & |\pm\rangle_1|+\rangle_2 |-\rangle_3
\xrightarrow{\text{block}_4}
{1 \over {2\sqrt {2}}}({p \over 2}({r_h} - {r_0}))^6(| {a_{11}^Rb_{11}^R} \rangle+ | {a_{12}^Rb_{12}^R} \rangle+| {a_{11}^Lb_{11}^L} \rangle\\&+| {a_{12}^Lb_{12}^L} \rangle
\pm(| {a_{21}^Rb_{21}^R} \rangle + | {a_{22}^Rb_{22}^R} \rangle + | {a_{22}^Lb_{22}^L} \rangle + | {a_{22}^Lb_{21}^L} \rangle))|ls,ls\rangle {|\pm\rangle_1}{|+\rangle_2}{|-\rangle_3},
\end{split}
\end{eqnarray}

\begin{eqnarray}              \label{eq59}
\begin{split}
({p \over 2}({r_h} - {r_0}))^6|\phi_S^\pm\rangle_{AB} |\psi_P^+\rangle_{AB} |\psi_T^-\rangle_{AB} & |\pm\rangle_1|+\rangle_2 |-\rangle_3
\xrightarrow{\text{block}_4}
{1 \over {2\sqrt {2}}}({p \over 2}({r_h} - {r_0}))^6(| {a_{11}^Rb_{11}^L} \rangle+ | {a_{12}^Rb_{12}^L} \rangle+| {a_{11}^Lb_{11}^R} \rangle\\&+ | {a_{12}^Lb_{12}^R}\rangle
\pm(| {a_{21}^Rb_{21}^L} \rangle + | {a_{21}^Lb_{21}^R} \rangle + | {a_{22}^Rb_{22}^L} \rangle + | {a_{22}^Lb_{21}^R} \rangle ))|ls,ls\rangle {|\pm\rangle_1}{|+\rangle_2}{|-\rangle _3},
\end{split}
\end{eqnarray}
\begin{eqnarray}              \label{eq60}
\begin{split}
({p \over 2}({r_h} - {r_0}))^6|\psi_S^\pm\rangle_{AB} |\phi_P^-\rangle_{AB} |\phi_T^+\rangle_{AB} & |\pm\rangle_1|+\rangle_2 |-\rangle_3
\xrightarrow{\text{block}_4}
{1 \over {2\sqrt {2}}}({p \over 2}({r_h} - {r_0}))^6(| {a_{11}^Rb_{21}^L} \rangle+| {a_{12}^Rb_{22}^L} \rangle+| {a_{12}^Lb_{22}^R} \rangle\\&+| {a_{11}^Lb_{21}^R} \rangle
\pm(| {a_{21}^Rb_{11}^L} \rangle+ | {a_{22}^Rb_{12}^L} \rangle+ | {a_{22}^Lb_{12}^R} \rangle + | {a_{21}^Lb_{11}^R} \rangle ))|ls,ls\rangle
 {|\pm\rangle_1}{|+\rangle_2}{|- \rangle _3},
\end{split}
\end{eqnarray}
\begin{eqnarray}              \label{eq61}
\begin{split}
({p \over 2}({r_h} - {r_0}))^6|\psi_S^\pm\rangle_{AB} |\phi_P^-\rangle_{AB} |\phi_T^-\rangle_{AB} & |\pm\rangle_1|+\rangle_2 |-\rangle_3
\xrightarrow{\text{block}_4}
{1 \over {2\sqrt {2}}}({p \over 2}({r_h} - {r_0}))^6(| {a_{11}^Rb_{21}^R} \rangle+| {a_{12}^Rb_{22}^R} \rangle+| {a_{12}^Lb_{22}^L} \rangle\\&+| {a_{11}^Lb_{21}^L} \rangle
\pm(| {a_{21}^Rb_{11}^R} \rangle+ | {a_{22}^Rb_{12}^R} \rangle+| {a_{22}^Lb_{12}^L} \rangle + | {a_{21}^Lb_{11}^L} \rangle ))|ls,ls\rangle
 {|\pm\rangle_1}{|+\rangle_2}{|-\rangle _3},
\end{split}
\end{eqnarray}

\begin{eqnarray}              \label{eq62}
\begin{split}
({p \over 2}({r_h} - {r_0}))^6|\psi_S^\pm\rangle_{AB} |\phi_P^-\rangle_{AB} |\psi_T^+\rangle_{AB} & |\pm\rangle_1|+\rangle_2 |-\rangle_3
\xrightarrow{\text{block}_4}
{1 \over {2\sqrt {2}}}({p \over 2}({r_h} - {r_0}))^6(| {a_{11}^Rb_{22}^L} \rangle-| {a_{12}^Rb_{21}^L} \rangle-| {a_{12}^Lb_{21}^R} \rangle\\&+| {a_{11}^Lb_{22}^R} \rangle
\pm(| {a_{21}^Rb_{12}^L} \rangle - | {a_{22}^Rb_{11}^L} \rangle- | {a_{22}^Lb_{11}^R} \rangle + | {a_{21}^Lb_{12}^R} \rangle ))|ls,ls\rangle
 {|\pm\rangle_1}{|+\rangle_2}{|-\rangle _3},
\end{split}
\end{eqnarray}
\begin{eqnarray}              \label{eq63}
\begin{split}
({p \over 2}({r_h} - {r_0}))^6|\psi_S^\pm\rangle_{AB} |\phi_P^-\rangle_{AB} |\psi_T^-\rangle_{AB} & |\pm\rangle_1|+\rangle_2 |-\rangle_3
\xrightarrow{\text{block}_4}
{1 \over {2\sqrt {2}}}({p \over 2}({r_h} - {r_0}))^6(| {a_{11}^Rb_{22}^R} \rangle-| {a_{12}^Rb_{21}^R} \rangle-| {a_{12}^Lb_{21}^L} \rangle\\&+| {a_{11}^Lb_{22}^L} \rangle
\pm(| {a_{21}^Rb_{12}^R} \rangle - | {a_{22}^Rb_{11}^R} \rangle- | {a_{22}^Lb_{11}^L} \rangle + | {a_{21}^Lb_{12}^L} \rangle ))|ls,ls\rangle
 {|\pm\rangle_1}{|+\rangle _2}{|-\rangle _3},
\end{split}
\end{eqnarray}
\begin{eqnarray}              \label{eq64}
\begin{split}
({p \over 2}({r_h} - {r_0}))^6|\psi_S^\pm\rangle_{AB} |\psi_P^+\rangle_{AB} |\phi_T^+\rangle_{AB} & |\pm\rangle_1|-\rangle_2 |+\rangle_3
\xrightarrow{\text{block}_4}
{1 \over {2\sqrt {2}}}({p \over 2}({r_h} - {r_0}))^6(| {a_{11}^Rb_{22}^R} \rangle-| {a_{12}^Rb_{21}^R} \rangle-| {a_{12}^Lb_{21}^L} \rangle\\&+| {a_{11}^Lb_{22}^L} \rangle
\pm(| {a_{21}^Rb_{12}^R}\rangle - |{a_{22}^Rb_{11}^R} \rangle- | {a_{22}^Lb_{11}^L} \rangle + | {a_{21}^Lb_{12}^L} \rangle))|ls,ls\rangle
 {|\pm\rangle_1}{|-\rangle_2}{| +\rangle _3},
\end{split}
\end{eqnarray}
\begin{eqnarray}              \label{eq65}
\begin{split}
({p \over 2}({r_h} - {r_0}))^6|\psi_S^\pm\rangle_{AB} |\psi_P^+\rangle_{AB} |\phi_T^-\rangle_{AB} & |\pm\rangle_1|-\rangle_2 |+\rangle_3
\xrightarrow{\text{block}_4}
{1 \over {2\sqrt {2}}}({p \over 2}({r_h} - {r_0}))^6(| {a_{11}^Rb_{22}^L} \rangle-| {a_{12}^Rb_{21}^L} \rangle-| {a_{12}^Lb_{21}^R} \rangle\\&+| {a_{11}^Lb_{22}^R} \rangle
\pm(| {a_{21}^Rb_{12}^L} \rangle - | {a_{22}^Rb_{11}^L} \rangle- | {a_{22}^Lb_{11}^R} \rangle + | {a_{21}^Lb_{12}^R} \rangle ))|ls,ls\rangle
 {|\pm\rangle _1}{|-\rangle _2}{| +\rangle _3},
\end{split}
\end{eqnarray}
\begin{eqnarray}              \label{eq66}
\begin{split}
({p \over 2}({r_h} - {r_0}))^6|\psi_S^\pm\rangle_{AB} |\psi_P^+\rangle_{AB} |\psi_T^+\rangle_{AB} & |\pm\rangle_1|-\rangle_2 |+\rangle_3
\xrightarrow{\text{block}_4}
{1 \over {2\sqrt {2}}}({p \over 2}({r_h} - {r_0}))^6(| {a_{11}^Rb_{21}^R} \rangle+| {a_{12}^Rb_{22}^R} \rangle+| {a_{12}^Lb_{22}^L} \rangle\\&+| {a_{11}^Lb_{21}^L} \rangle
\pm(| {a_{21}^Rb_{11}^R} \rangle+ | {a_{22}^Rb_{12}^R} \rangle+| {a_{22}^Lb_{12}^L} \rangle + | {a_{21}^Lb_{11}^L} \rangle ))|ls,ls\rangle
 {|   \pm  \rangle _1}{|  -  \rangle _2}{| +\rangle _3},
\end{split}
\end{eqnarray}

\begin{eqnarray}              \label{eq67}
\begin{split}
({p \over 2}({r_h} - {r_0}))^6|\psi_S^\pm\rangle_{AB} |\psi_P^+\rangle_{AB} |\psi_T^-\rangle_{AB} & |\pm\rangle_1|-\rangle_2 |+\rangle_3
\xrightarrow{\text{block}_4}
{1 \over {2\sqrt {2}}}({p \over 2}({r_h} - {r_0}))^6(| {a_{11}^Rb_{21}^L} \rangle+| {a_{12}^Rb_{22}^L} \rangle+| {a_{12}^Lb_{22}^R} \rangle\\&+| {a_{11}^Lb_{21}^R} \rangle
\pm(| {a_{21}^Rb_{11}^L} \rangle+ | {a_{22}^Rb_{12}^L} \rangle+ | {a_{22}^Lb_{12}^R} \rangle + | {a_{21}^Lb_{11}^R} \rangle ))|ls,ls\rangle
 {|   \pm  \rangle _1}{|  -  \rangle _2}{|+ \rangle _3},
\end{split}
\end{eqnarray}
\begin{eqnarray}              \label{eq68}
\begin{split}
({p \over 2}({r_h} - {r_0}))^6|\phi_S^\pm\rangle_{AB} |\phi_P^-\rangle_{AB} |\phi_T^+\rangle_{AB} & |\pm\rangle_1|-\rangle_2 |+\rangle_3
\xrightarrow{\text{block}_4}
{1 \over {2\sqrt {2}}}({p\over 2}({r_h} - {r_0}))^6 (| {a_{11}^Rb_{11}^L} \rangle- | {a_{12}^Rb_{12}^L} \rangle+| {a_{11}^Lb_{11}^R} \rangle\\&-| {a_{12}^Lb_{12}^R}\rangle
\pm(| {a_{21}^Rb_{21}^L} \rangle + | {a_{21}^Lb_{21}^R} \rangle- | {a_{22}^Rb_{22}^L} \rangle-| {a_{22}^Lb_{21}^R} \rangle ))|ls,ls\rangle
 {|\pm\rangle _1}{|- \rangle _2}{| + \rangle _3},
\end{split}
\end{eqnarray}
\begin{eqnarray}              \label{eq69}
\begin{split}
({p \over 2}({r_h} - {r_0}))^6|\phi_S^\pm\rangle_{AB} |\phi_P^-\rangle_{AB} |\phi_T^-\rangle_{AB} & |\pm\rangle_1 |-\rangle_2 |+\rangle_3
\xrightarrow{\text{block}_4}
{1 \over {2\sqrt {2}}}({p \over 2}({r_h} - {r_0}))^6(| {a_{11}^Rb_{11}^R} \rangle+ | {a_{12}^Rb_{12}^R} \rangle+| {a_{11}^Lb_{11}^L} \rangle\\&+| {a_{12}^Lb_{12}^L} \rangle
\pm(| {a_{21}^Rb_{21}^R} \rangle- | {a_{22}^Rb_{22}^R} \rangle-| {a_{22}^Lb_{22}^L} \rangle + | {a_{21}^Lb_{21}^L} \rangle ))|ls,ls\rangle
 {|   \pm  \rangle _1}{|  -  \rangle _2}{| +\rangle _3},
\end{split}
\end{eqnarray}
\begin{eqnarray}              \label{eq70}
\begin{split}
({p \over 2}({r_h} - {r_0}))^6|\phi_S^\pm\rangle_{AB} |\phi_P^-\rangle_{AB} |\psi_T^+\rangle_{AB} & |\pm\rangle_1|-\rangle_2 |+\rangle_3
\xrightarrow{\text{block}_4}
{1 \over {2\sqrt {2}}}({p \over 2}({r_h} - {r_0}))^6(| {a_{11}^Rb_{12}^L} \rangle- | {a_{12}^Rb_{11}^L} \rangle-| {a_{11}^Lb_{12}^R} \rangle\\&+| {a_{12}^Lb_{11}^R}\rangle
\pm(| {a_{21}^Rb_{22}^L} \rangle- | {a_{22}^Rb_{21}^L} \rangle+| {a_{22}^Lb_{21}^R} \rangle -| {a_{22}^Lb_{22}^R} \rangle))|ls,ls\rangle |\pm\rangle_1|-\rangle_2 |+\rangle_3,
\end{split}
\end{eqnarray}

\begin{eqnarray}              \label{eq71}
\begin{split}
({p \over 2}({r_h} - {r_0}))^6|\phi_S^\pm\rangle_{AB} |\phi_P^-\rangle_{AB} |\psi_T^-\rangle_{AB} & |\pm\rangle_1|-\rangle_2 |+\rangle_3
\xrightarrow{\text{block}_4}
{1 \over {2\sqrt {2}}}({p \over 2}({r_h} - {r_0}))^6(| {a_{11}^Rb_{12}^R} \rangle-| {a_{12}^Rb_{11}^R}\rangle-| {a_{11}^Lb_{12}^L} \rangle\\&+| {a_{12}^Lb_{11}^L} \rangle
\pm(| {a_{21}^Rb_{22}^R} \rangle -| {a_{22}^Rb_{21}^R} \rangle+| {a_{22}^Lb_{21}^L} \rangle -| {a_{22}^Lb_{22}^L} \rangle ))|ls,ls\rangle
 {| \pm \rangle _1}{|-\rangle_2}{| +\rangle _3},
\end{split}
\end{eqnarray}

\begin{eqnarray}              \label{eq48}
\begin{split}
({p \over 2}({r_h} - {r_0}))^6|\psi_S^\pm\rangle_{AB} |\phi_P^+\rangle_{AB} |\phi_T^+\rangle_{AB} & |\pm\rangle_1|-\rangle_2 |-\rangle_3
\xrightarrow{\text{block}_4}
{1 \over {2\sqrt {2}}}({p \over 2}({r_h} - {r_0}))^6(| {a_{11}^Rb_{21}^R} \rangle+| {a_{12}^Rb_{22}^R} \rangle+| {a_{12}^Lb_{22}^L} \rangle\\&+| {a_{11}^Lb_{21}^L} \rangle
\pm(| {a_{21}^Rb_{11}^R} \rangle+ | {a_{22}^Rb_{12}^R} \rangle+| {a_{22}^Lb_{12}^L} \rangle + |{a_{21}^Lb_{11}^L} \rangle ))|ls,ls\rangle
 {|\pm\rangle_1}{|-\rangle_2}{|-\rangle _3},
\end{split}
\end{eqnarray}
\begin{eqnarray}              \label{eq49}
\begin{split}
({p \over 2}({r_h} - {r_0}))^6|\psi_S^\pm\rangle_{AB} |\phi_P^+\rangle_{AB} |\phi_T^-\rangle_{AB} & |\pm\rangle_1|-\rangle_2 |-\rangle_3
\xrightarrow{\text{block}_4}
{1 \over {2\sqrt {2}}}({p \over 2}({r_h} - {r_0}))^6(| {a_{11}^Rb_{21}^L} \rangle+| {a_{12}^Rb_{22}^L} \rangle+| {a_{12}^Lb_{22}^R} \rangle\\&+| {a_{11}^Lb_{21}^R} \rangle
\pm(| {a_{21}^Rb_{11}^L} \rangle+ | {a_{22}^Rb_{12}^L} \rangle+ | {a_{22}^Lb_{12}^R} \rangle + | {a_{21}^Lb_{11}^R} \rangle))|ls,ls\rangle
 {|\pm \rangle _1}{|- \rangle _2}{| -\rangle _3},
\end{split}
\end{eqnarray}

\begin{eqnarray}              \label{eq50}
\begin{split}
({p \over 2}({r_h} - {r_0}))^6|\psi_S^\pm\rangle_{AB} |\phi_P^+\rangle_{AB} |\psi_T^+\rangle_{AB} & |\pm\rangle_1|-\rangle_2 |-\rangle_3
\xrightarrow{\text{block}_4}
{1 \over {2\sqrt {2}}}({p \over 2}({r_h} - {r_0}))^6(| {a_{11}^Rb_{22}^R} \rangle+| {a_{12}^Rb_{21}^R} \rangle-| {a_{12}^Lb_{21}^L} \rangle\\&-| {a_{11}^Lb_{22}^L} \rangle
\pm(| {a_{21}^Rb_{12}^R} \rangle+| {a_{22}^Rb_{11}^R} \rangle- | {a_{22}^Lb_{11}^L} \rangle- | {a_{21}^Lb_{12}^L} \rangle ))|ls,ls\rangle
 {|\pm\rangle _1}{|-\rangle _2}{| - \rangle _3},
\end{split}
\end{eqnarray}
\begin{eqnarray}              \label{eq51}
\begin{split}
({p \over 2}({r_h} - {r_0}))^6|\psi_S^\pm\rangle_{AB} |\phi_P^+\rangle_{AB} |\psi_T^-\rangle_{AB} & |\pm\rangle_1|-\rangle_2 |-\rangle_3
\xrightarrow{\text{block}_4}
{1 \over {2\sqrt {2}}}({p \over 2}({r_h} - {r_0}))^6(| {a_{11}^Rb_{22}^L} \rangle+| {a_{12}^Rb_{21}^L} \rangle-| {a_{12}^Lb_{21}^R} \rangle\\&-| {a_{11}^Lb_{22}^R} \rangle
\pm(| {a_{21}^Rb_{12}^L} \rangle+ | {a_{22}^Rb_{11}^L} \rangle- | {a_{22}^Lb_{11}^R} \rangle- | {a_{21}^Lb_{12}^R} \rangle ))|ls,ls\rangle
 {|\pm\rangle _1}{|-\rangle _2}{|-\rangle _3},
\end{split}
\end{eqnarray}
\begin{eqnarray}              \label{eq52}
\begin{split}
({p \over 2}({r_h} - {r_0}))^6|\phi_S^\pm\rangle_{AB} |\psi_P^-\rangle_{AB} |\phi_T^+\rangle_{AB} & |\pm\rangle_1|-\rangle_2 |-\rangle_3
\xrightarrow{\text{block}_4}
{1 \over {2\sqrt {2}}}({p \over 2}({r_h} - {r_0}))^6(| {a_{11}^Rb_{12}^L} \rangle+ | {a_{12}^Rb_{11}^L} \rangle-| {a_{11}^Lb_{12}^R} \rangle\\&-| {a_{12}^Lb_{11}^R}\rangle
\pm(| {a_{21}^Rb_{22}^L} \rangle + | {a_{22}^Rb_{21}^L} \rangle-| {a_{22}^Lb_{21}^R} \rangle -| {a_{22}^Lb_{22}^R} \rangle ))|ls,ls\rangle
 {|\pm\rangle _1}{|-\rangle_2}{|-\rangle _3},
\end{split}
\end{eqnarray}
\begin{eqnarray}              \label{eq53}
\begin{split}
({p \over 2}({r_h} - {r_0}))^6|\phi_S^\pm\rangle_{AB} |\psi_P^-\rangle_{AB} |\phi_T^-\rangle_{AB} & |\pm\rangle_1|-\rangle_2 |-\rangle_3
\xrightarrow{\text{block}_4}
{1 \over {2\sqrt {2}}}({p \over 2}({r_h} - {r_0}))^6(|{a_{11}^Rb_{12}^R} \rangle+ |{a_{12}^Rb_{11}^R}\rangle-|{a_{11}^Lb_{12}^L} \rangle\\&-|{a_{12}^Lb_{11}^L} \rangle
\pm(|{a_{21}^Rb_{22}^R} \rangle + |{a_{22}^Rb_{21}^R} \rangle- |{a_{22}^Lb_{21}^L} \rangle -|{a_{22}^Lb_{22}^L} \rangle))|ls,ls\rangle
 {|\pm\rangle_1}{|-\rangle_2}{|-\rangle _3},
\end{split}
\end{eqnarray}
\begin{eqnarray}              \label{eq54}
\begin{split}
({p \over 2}({r_h} - {r_0}))^6|\phi_S^\pm\rangle_{AB} |\psi_P^-\rangle_{AB} |\psi_T^+\rangle_{AB} & |\pm\rangle_1|-\rangle_2 |-\rangle_3
\xrightarrow{\text{block}_4}
{1 \over {2\sqrt {2}}}({p \over 2}({r_h} - {r_0}))^6(| {a_{11}^Rb_{11}^L} \rangle- | {a_{12}^Rb_{12}^L}\rangle-| {a_{11}^Lb_{11}^R} \rangle\\&+| {a_{12}^Lb_{12}^R} \rangle
\pm(| {a_{21}^Rb_{21}^L} \rangle- | {a_{22}^Rb_{22}^L} \rangle- | {a_{21}^Lb_{21}^R} \rangle+| {a_{22}^Lb_{22}^R} \rangle))|ls,ls\rangle
 {|   \pm  \rangle_1}{|- \rangle_2}{|-\rangle _3},
\end{split}
\end{eqnarray}
\begin{eqnarray}              \label{eq55}
\begin{split}
({p \over 2}({r_h} - {r_0}))^6|\phi_S^\pm\rangle_{AB} |\psi_P^-\rangle_{AB} |\psi_T^-\rangle_{AB} & |\pm\rangle_1|-\rangle_2 |-\rangle_3
\xrightarrow{\text{block}_4}
{1 \over {2\sqrt {2}}}({p \over 2}({r_h} - {r_0}))^6(|a_{11}^R b_{11}^R\rangle- |a_{12}^R b_{12}^R\rangle-|a_{11}^Lb_{11}^L\rangle\\&+|a_{12}^L b_{12}^L\rangle
\pm(|a_{21}^R b_{21}^R\rangle- |a_{22}^R b_{22}^R\rangle- |a_{21}^L b_{21}^L\rangle+|a_{22}^Lb_{22}^L\rangle))|ls,ls\rangle|\pm\rangle_1|-\rangle_2|-\rangle_3.
\end{split}
\end{eqnarray}
Here the Pockel cell PC, which is used to flip the polarization states of the incident photons at a definite time, has been achieved with efficiency more than 99\%.\upcite{PC}

Based on above discussion, one can see that  the 64 hyper-Bell states can be divided into 8 distinct groups according to the three QD-spin states, and the 8 hyper-Bell states in each group can be further distinguished from each other by the clicks of the single-photon detectors. Therefore, with the scheme depicted by Figure \ref{local}, the 64 hyper-Bell states in polarization, spatial, and time-bin DOFs can be completely distinguished.  The correspondences between the hyper-Bell states, the spins of the three QDs, and the detection results are shown in detail in Table \ref{table1}.
For instance, if the outcomes of the three QDs are $|-\rangle _1 |+\rangle _2 |+\rangle _3$ and the detection signature is one of
\{$a_{11}^{R}b_{22}^{R}$, $a_{12}^{R}b_{21}^{R}$, $a_{21}^{R}b_{12}^{R}$, $a_{22}^{R}b_{11}^{R}$, $a_{11}^{L}b_{22}^{L}$, $a_{12}^{L}b_{21}^{L}$,
$a_{21}^{L}b_{12}^{L}$, $a_{22}^{L}b_{11}^{L}$\},  we can deduce that the initial hyper-Bell state is $|{\phi^-_S}\rangle|{\psi^+_P}\rangle|{\phi^-_T}\rangle$.

\begin{table}[htb] 
\centering \caption{The correspondences between the 64 hyper-Bell states and the outcomes of the three QDs and two incident photons.}
\begin{tabular}{ccccccc}
\hline
States & QDs & Single-photon detectors \\
\hline

$|\phi^+_S\rangle_{AB}|\phi^+_P\rangle_{AB}|\phi^{\pm}_T\rangle_{AB}$&\multirow{4}{*}{$|+\rangle_1|+\rangle_2|+\rangle_3$}
&$a_{11}^{R}b_{11}^{R/L}$, ${a_{12}^{R}}{b_{12}^{R/L}}$, ${a_{21}^{R}}{b_{21}^{R/L}}$, ${a_{22}^{R}}{b_{22}^{R/L}}$, ${a_{11}^{L}}{b_{11}^{L/R}}$, ${a_{12}^{L}}{b_{12}^{L/R}}$, ${a_{21}^{L}}{b_{21}^{L/R}}$, ${a_{22}^{L}}{b_{22}^{L/R}}$\\

$|\phi^+_S\rangle_{AB}|\psi^-_P\rangle_{AB}|\psi^{\pm}_T\rangle_{AB}$&
&${a_{11}^{R}}{b_{21}^{R/L}}$, ${a_{12}^{R}}{b_{22}^{R/L}}$, ${a_{21}^{R}}{b_{11}^{R/L}}$, ${a_{22}^{R}}{b_{12}^{R/L}}$, ${a_{11}^{L}}{b_{21}^{L/R}}$, ${a_{12}^{L}}{b_{22}^{L/R}}$, ${a_{21}^{L}}{b_{11}^{L/R}}$, ${a_{22}^{L}}{b_{12}^{L/R}}$\\

$|\phi^+_S\rangle_{AB}|\phi^+_P\rangle_{AB}|\psi^{\pm}_T\rangle_{AB}$&
&${a_{11}^{R}}{b_{12}^{R/L}}$, ${a_{12}^{R}}{b_{11}^{R/L}}$, ${a_{21}^{R}}{b_{22}^{R/L}}$, ${a_{22}^{R}}{b_{21}^{R/L}}$, ${a_{11}^{L}}{b_{12}^{L/R}}$, ${a_{12}^{L}}{b_{11}^{L/R}}$, ${a_{21}^{L}}{b_{22}^{L/R}}$, ${a_{22}^{L}}{b_{21}^{L/R}}$\\

$|\phi^+_S\rangle_{AB}|\psi^-_P\rangle_{AB}|\phi^{\pm}_T\rangle_{AB}$&
&${a_{11}^{R}}{b_{22}^{R/L}}$, ${a_{12}^{R}}{b_{21}^{R/L}}$, ${a_{21}^{R}}{b_{12}^{R/L}}$, ${a_{22}^{R}}{b_{11}^{R/L}}$, ${a_{11}^{L}}{b_{22}^{L/R}}$, ${a_{12}^{L}}{b_{21}^{L/R}}$, ${a_{21}^{L}}{b_{12}^{L/R}}$, ${a_{22}^{L}}{b_{11}^{L/R}}$\\
\noalign{\smallskip}\hline\noalign{\smallskip}

$|\phi^+_S\rangle_{AB}|\phi^-_P\rangle_{AB}|\psi^{\pm}_T\rangle_{AB}$&\multirow{4}{*}{$|+\rangle_1|+\rangle_2|-\rangle_3$}
&${a_{11}^{R}}{b_{11}^{R/L}}$, ${a_{12}^{R}}{b_{12}^{R/L}}$, ${a_{21}^{R}}{b_{21}^{R/L}}$, ${a_{22}^{R}}{b_{22}^{R/L}}$, ${a_{11}^{L}}{b_{11}^{L/R}}$, ${a_{12}^{L}}{b_{12}^{L/R}}$, ${a_{21}^{L}}{b_{21}^{L/R}}$, ${a_{22}^{L}}{b_{22}^{L/R}}$\\

$|\phi^+_S\rangle_{AB}|\psi^+_P\rangle_{AB}|\phi^{\pm}_T\rangle_{AB}$&
&${a_{11}^{R}}{b_{21}^{R/L}}$, ${a_{12}^{R}}{b_{22}^{R/L}}$, ${a_{21}^{R}}{b_{11}^{R/L}}$, ${a_{22}^{R}}{b_{12}^{R/L}}$,
${a_{11}^{L}}{b_{21}^{L/R}}$, ${a_{12}^{L}}{b_{22}^{L/R}}$, ${a_{21}^{L}}{b_{11}^{L/R}}$, ${a_{22}^{L}}{b_{12}^{L/R}}$\\

$|\phi^+_S\rangle_{AB}|\phi^-_P\rangle_{AB}|\phi^{\pm}_T\rangle_{AB}$&
&${a_{11}^{R}}{b_{12}^{R/L}}$, ${a_{12}^{R}}{b_{11}^{R/L}}$, ${a_{21}^{R}}{b_{22}^{R/L}}$, ${a_{22}^{R}}{b_{21}^{R/L}}$,
${a_{11}^{L}}{b_{12}^{L/R}}$, ${a_{12}^{L}}{b_{11}^{L/R}}$, ${a_{21}^{L}}{b_{22}^{L/R}}$, ${a_{22}^{L}}{b_{21}^{L/R}}$\\

$|\phi^+_S\rangle_{AB}|\psi^+_P\rangle_{AB}|\psi^{\pm}_T\rangle_{AB}$&
&${a_{11}^{R}}{b_{22}^{R/L}}$, ${a_{12}^{R}}{b_{21}^{R/L}}$, ${a_{21}^{R}}{b_{12}^{R/L}}$, ${a_{22}^{R}}{b_{11}^{R/L}}$,
${a_{11}^{L}}{b_{22}^{L/R}}$, ${a_{12}^{L}}{b_{21}^{L/R}}$, ${a_{21}^{L}}{b_{12}^{L/R}}$, ${a_{22}^{L}}{b_{11}^{L/R}}$\\
\noalign{\smallskip}\hline\noalign{\smallskip}

$|\psi^+_S\rangle_{AB}|\psi^+_P\rangle_{AB}|\phi^{\pm}_T\rangle_{AB}$&\multirow{4}{*}{$|+\rangle_1|-\rangle_2|+\rangle_3$}
&${a_{11}^{R}}{b_{11}^{R/L}}$, ${a_{12}^{R}}{b_{12}^{R/L}}$, ${a_{21}^{R}}{b_{21}^{R/L}}$, ${a_{22}^{R}}{b_{22}^{R/L}}$, ${a_{11}^{L}}{b_{11}^{L/R}}$, ${a_{12}^{L}}{b_{12}^{L/R}}$, ${a_{21}^{L}}{b_{21}^{L/R}}$, ${a_{22}^{L}}{b_{22}^{L/R}}$\\

$|\psi^+_S\rangle_{AB}|\phi^-_P\rangle_{AB}|\psi^{\pm}_T\rangle_{AB}$&
&${a_{11}^{R}}{b_{21}^{R/L}}$, ${a_{12}^{R}}{b_{22}^{R/L}}$, ${a_{21}^{R}}{b_{11}^{R/L}}$, ${a_{22}^{R}}{b_{12}^{R/L}}$, ${a_{11}^{L}}{b_{21}^{L/R}}$, ${a_{12}^{L}}{b_{22}^{L/R}}$, ${a_{21}^{L}}{b_{11}^{L/R}}$, ${a_{22}^{L}}{b_{12}^{L/R}}$\\

$|\psi^+_S\rangle_{AB}|\psi^+_P\rangle_{AB}|\psi^{\pm}_T\rangle_{AB}$&
&${a_{11}^{R}}{b_{12}^{R/L}}$, ${a_{12}^{R}}{b_{11}^{R/L}}$, ${a_{21}^{R}}{b_{22}^{R/L}}$, ${a_{22}^{R}}{b_{21}^{R/L}}$, ${a_{11}^{L}}{b_{12}^{L/R}}$, ${a_{12}^{L}}{b_{11}^{L/R}}$, ${a_{21}^{L}}{b_{22}^{L/R}}$, ${a_{22}^{L}}{b_{21}^{L/R}}$\\

$|\psi^+_S\rangle_{AB}|\phi^-_P\rangle_{AB}|\phi^{\pm}_T\rangle_{AB}$&
&${a_{11}^{R}}{b_{22}^{R/L}}$, ${a_{12}^{R}}{b_{21}^{R/L}}$, ${a_{21}^{R}}{b_{12}^{R/L}}$, ${a_{22}^{R}}{b_{11}^{R/L}}$, ${a_{11}^{L}}{b_{22}^{L/R}}$, ${a_{12}^{L}}{b_{21}^{L/R}}$, ${a_{21}^{L}}{b_{12}^{L/R}}$, ${a_{22}^{L}}{b_{11}^{L/R}}$\\
\noalign{\smallskip}\hline\noalign{\smallskip}

$|\psi^+_S\rangle_{AB}|\psi^-_P\rangle_{AB}|\psi^{\pm}_T\rangle_{AB}$&\multirow{4}{*}{$|+\rangle_1|-\rangle_2|-\rangle_3$}
&${a_{11}^{R}}{b_{11}^{R/L}}$, ${a_{12}^{R}}{b_{12}^{R/L}}$, ${a_{21}^{R}}{b_{21}^{R/L}}$, ${a_{22}^{R}}{b_{22}^{R/L}}$, ${a_{11}^{L}}{b_{11}^{L/R}}$, ${a_{12}^{L}}{b_{12}^{L/R}}$, ${a_{21}^{L}}{b_{21}^{L/R}}$, ${a_{22}^{L}}{b_{22}^{L/R}}$\\

$|\psi^+_S\rangle_{AB}|\phi^+_P\rangle_{AB}|\phi^{\pm}_T\rangle_{AB}$&
&${a_{11}^{R}}{b_{21}^{R/L}}$, ${a_{12}^{R}}{b_{22}^{R/L}}$, ${a_{21}^{R}}{b_{11}^{R/L}}$, ${a_{22}^{R}}{b_{12}^{R/L}}$, ${a_{11}^{L}}{b_{21}^{L/R}}$, ${a_{12}^{L}}{b_{22}^{L/R}}$, ${a_{21}^{L}}{b_{11}^{L/R}}$, ${a_{22}^{L}}{b_{12}^{L/R}}$\\

$|\psi^+_S\rangle_{AB}|\psi^-_P\rangle_{AB}|\phi^{\pm}_T\rangle_{AB}$&
&${a_{11}^{R}}{b_{12}^{R/L}}$, ${a_{12}^{R}}{b_{11}^{R/L}}$, ${a_{21}^{R}}{b_{22}^{R/L}}$, ${a_{22}^{R}}{b_{21}^{R/L}}$, ${a_{11}^{L}}{b_{12}^{L/R}}$, ${a_{12}^{L}}{b_{11}^{L/R}}$, ${a_{21}^{L}}{b_{22}^{L/R}}$, ${a_{22}^{L}}{b_{21}^{L/R}}$\\

$|\psi^+_S\rangle_{AB}|\phi^+_P\rangle_{AB}|\psi^{\pm}_T\rangle_{AB}$&
&${a_{11}^{R}}{b_{22}^{R/L}}$, ${a_{12}^{R}}{b_{21}^{R/L}}$, ${a_{21}^{R}}{b_{12}^{R/L}}$, ${a_{22}^{R}}{b_{11}^{R/L}}$, ${a_{11}^{L}}{b_{22}^{L/R}}$, ${a_{12}^{L}}{b_{21}^{L/R}}$, ${a_{21}^{L}}{b_{12}^{L/R}}$, ${a_{22}^{L}}{b_{11}^{L/R}}$\\
\noalign{\smallskip}\hline\noalign{\smallskip}

$|\phi^-_S\rangle_{AB}|\phi^-_P\rangle_{AB}|\phi^{\pm}_T\rangle_{AB}$&\multirow{4}{*}{$|-\rangle_1|+\rangle_2|+\rangle_3$}
&${a_{11}^{R}}{b_{11}^{R/L}}$, ${a_{12}^{R}}{b_{12}^{R/L}}$, ${a_{21}^{R}}{b_{21}^{R/L}}$, ${a_{22}^{R}}{b_{22}^{R/L}}$, ${a_{11}^{L}}{b_{11}^{L/R}}$, ${a_{12}^{L}}{b_{12}^{L/R}}$, ${a_{21}^{L}}{b_{21}^{L/R}}$, ${a_{22}^{L}}{b_{22}^{L/R}}$\\

$|\phi^-_S\rangle_{AB}|\psi^+_P\rangle_{AB}|\psi^{\pm}_T\rangle_{AB}$&
&${a_{11}^{R}}{b_{21}^{R/L}}$, ${a_{12}^{R}}{b_{22}^{R/L}}$, ${a_{21}^{R}}{b_{11}^{R/L}}$, ${a_{22}^{R}}{b_{12}^{R/L}}$, ${a_{11}^{L}}{b_{21}^{L/R}}$, ${a_{12}^{L}}{b_{22}^{L/R}}$, ${a_{21}^{L}}{b_{11}^{L/R}}$, ${a_{22}^{L}}{b_{12}^{L/R}}$\\

$|\phi^-_S\rangle_{AB}|\phi^-_P\rangle_{AB}|\psi^{\pm}_T\rangle_{AB}$&
&${a_{11}^{R}}{b_{12}^{R/L}}$, ${a_{12}^{R}}{b_{11}^{R/L}}$, ${a_{21}^{R}}{b_{22}^{R/L}}$,
${a_ {22}^{R}}{b_{21}^{R/L}}$, ${a_{11}^{L}}{b_{12}^{L/R}}$, ${a_{12}^{L}}{b_{11}^{L/R}}$, ${a_{21}^{L}}{b_{22}^{L/R}}$, ${a_{22}^{L}}{b_{21}^{L/R}}$\\

$|\phi^-_S\rangle_{AB}|\psi^+_P\rangle_{AB}|\phi^{\pm}_T\rangle_{AB}$&
&${a_{11}^{R}}{b_{22}^{R/L}}$, ${a_{12}^{R}}{b_{21}^{R/L}}$, ${a_{21}^{R}}{b_{12}^{R/L}}$, ${a_{22}^{R}}{b_{11}^{R/L}}$, ${a_{11}^{L}}{b_{22}^{L/R}}$, ${a_{12}^{L}}{b_{21}^{L/R}}$, ${a_{21}^{L}}{b_{12}^{L/R}}$, ${a_{22}^{L}}{b_{11}^{L/R}}$\\
\noalign{\smallskip}\hline\noalign{\smallskip}

$|\phi^-_S\rangle_{AB}|\phi^+_P\rangle_{AB}|\psi^{\pm}_T\rangle_{AB}$&\multirow{4}{*}{$|-\rangle_1|+\rangle_2|-\rangle_3$}
&${a_{11}^{R}}{b_{11}^{R/L}}$, ${a_{12}^{R}}{b_{12}^{R/L}}$, ${a_{21}^{R}}{b_{21}^{R/L}}$, ${a_{22}^{R}}{b_{22}^{R/L}}$, ${a_{11}^{L}}{b_{11}^{L/R}}$, ${a_{12}^{L}}{b_{12}^{L/R}}$, ${a_{21}^{L}}{b_{21}^{L/R}}$, ${a_{22}^{L}}{b_{22}^{L/R}}$\\

$|\phi^-_S\rangle_{AB}|\psi^-_P\rangle_{AB}|\phi^{\pm}_T\rangle_{AB}$&
&${a_{11}^{R}}{b_{21}^{R/L}}$, ${a_{12}^{R}}{b_{22}^{R/L}}$, ${a_{21}^{R}}{b_{11}^{R/L}}$, ${a_{22}^{R}}{b_{12}^{R/L}}$, ${a_{11}^{L}}{b_{21}^{L/R}}$, ${a_{12}^{L}}{b_{22}^{L/R}}$, ${a_{21}^{L}}{b_{11}^{L/R}}$, ${a_{22}^{L}}{b_{12}^{L/R}}$\\

$|\phi^-_S\rangle_{AB}|\psi^-_P\rangle_{AB}|\psi^{\pm}_T\rangle_{AB}$&
&${a_{11}^{R}}{b_{22}^{R/L}}$, ${a_{12}^{R}}{b_{21}^{R/L}}$, ${a_{21}^{R}}{b_{12}^{R/L}}$, ${a_{22}^{R}}{b_{11}^{R/L}}$, ${a_{11}^{L}}{b_{22}^{L/R}}$, ${a_{12}^{L}}{b_{21}^{L/R}}$, ${a_{21}^{L}}{b_{12}^{L/R}}$, ${a_{22}^{L}}{b_{11}^{L/R}}$\\

$|\phi^-_S\rangle_{AB}|\phi^+_P\rangle_{AB}|\phi^{\pm}_T\rangle_{AB}$&
&${a_{11}^{R}}{b_{12}^{R/L}}$, ${a_{12}^{R}}{b_{11}^{R/L}}$, ${a_{21}^{R}}{b_{22}^{R/L}}$, ${a_{22}^{R}}{b_{21}^{R/L}}$, ${a_{11}^{L}}{b_{12}^{L/R}}$, ${a_{12}^{L}}{b_{11}^{L/R}}$, ${a_{21}^{L}}{b_{22}^{L/R}}$, ${a_{22}^{L}}{b_{21}^{L/R}}$\\
\noalign{\smallskip}\hline\noalign{\smallskip}

$|\psi^-_S\rangle_{AB}|\psi^-_P\rangle_{AB}|\phi^{\pm}_T\rangle_{AB}$&\multirow{4}{*}{$|-\rangle_1|-\rangle_2|+\rangle_3$}
&${a_{11}^{R}}{b_{11}^{R/L}}$, ${a_{12}^{R}}{b_{12}^{R/L}}$, ${a_{21}^{R}}{b_{21}^{R/L}}$, ${a_{22}^{R}}{b_{22}^{R/L}}$, ${a_{11}^{L}}{b_{11}^{L/R}}$, ${a_{12}^{L}}{b_{12}^{L/R}}$, ${a_{21}^{L}}{b_{21}^{L/R}}$, ${a_{22}^{L}}{b_{22}^{L/R}}$\\

$|\psi^-_S\rangle_{AB}|\phi^+_P\rangle_{AB}|\psi^{\pm}_T\rangle_{AB}$&
&${a_{11}^{R}}{b_{21}^{R/L}}$, ${a_{12}^{R}}{b_{22}^{R/L}}$, ${a_{21}^{R}}{b_{11}^{R/L}}$, ${a_{22}^{R}}{b_{12}^{R/L}}$, ${a_{11}^{L}}{b_{21}^{L/R}}$, ${a_{12}^{L}}{b_{22}^{L/R}}$, ${a_{21}^{L}}{b_{11}^{L/R}}$, ${a_{22}^{L}}{b_{12}^{L/R}}$\\

$|\psi^-_S\rangle_{AB}|\psi^-_P\rangle_{AB}|\psi^{\pm}_T\rangle_{AB}$&
&${a_{11}^{R}}{b_{12}^{R/L}}$, ${a_{12}^{R}}{b_{11}^{R/L}}$, ${a_{21}^{R}}{b_{22}^{R/L}}$, ${a_{22}^{R}}{b_{21}^{R/L}}$, ${a_{11}^{L}}{b_{12}^{L/R}}$, ${a_{12}^{L}}{b_{11}^{L/R}}$, ${a_{21}^{L}}{b_{22}^{L/R}}$, ${a_{22}^{L}}{b_{21}^{L/R}}$\\

$|\psi^-_S\rangle_{AB}|\phi^+_P\rangle_{AB}|\phi^{\pm}_T\rangle_{AB}$&
&${a_{11}^{R}}{b_{22}^{R/L}}$, ${a_{12}^{R}}{b_{21}^{R/L}}$, ${a_{21}^{R}}{b_{12}^{R/L}}$, ${a_{22}^{R}}{b_{11}^{R/L}}$, ${a_{11}^{L}}{b_{22}^{L/R}}$, ${a_{12}^{L}}{b_{21}^{L/R}}$, ${a_{21}^{L}}{b_{12}^{L/R}}$, ${a_{22}^{L}}{b_{11}^{L/R}}$\\
\noalign{\smallskip}\hline\noalign{\smallskip}

$|\psi^-_S\rangle_{AB}|\psi^+_P\rangle_{AB}|\psi^{\pm}_T\rangle_{AB}$&\multirow{4}{*}{$|-\rangle_1|-\rangle_2|-\rangle_3$}
&${a_{11}^{R}}{b_{11}^{R/L}}$, ${a_{12}^{R}}{b_{12}^{R/L}}$, ${a_{21}^{R}}{b_{21}^{R/L}}$, ${a_{22}^{R}}{b_{22}^{R/L}}$, ${a_{11}^{L}}{b_{11}^{L/R}}$, ${a_{12}^{L}}{b_{12}^{L/R}}$, ${a_{21}^{L}}{b_{21}^{L/R}}$, ${a_{22}^{L}}{b_{22}^{L/R}}$\\

$|\psi^-_S\rangle_{AB}|\phi^-_P\rangle_{AB}|\phi^{\pm}_T\rangle_{AB}$&
&${a_{11}^{R}}{b_{21}^{R/L}}$, ${a_{12}^{R}}{b_{22}^{R/L}}$, ${a_{21}^{R}}{b_{11}^{R/L}}$, ${a_{22}^{R}}{b_{12}^{R/L}}$, ${a_{11}^{L}}{b_{21}^{L/R}}$, ${a_{12}^{L}}{b_{22}^{L/R}}$, ${a_{21}^{L}}{b_{11}^{L/R}}$, ${a_{22}^{L}}{b_{12}^{L/R}}$\\

$|\psi^-_S\rangle_{AB}|\psi^+_P\rangle_{AB}|\phi^{\pm}_T\rangle_{AB}$&
&${a_{11}^{R}}{b_{12}^{R/L}}$, ${a_{12}^{R}}{b_{11}^{R/L}}$, ${a_{21}^{R}}{b_{22}^{R/L}}$, ${a_{22}^{R}}{b_{21}^{R/L}}$, ${a_{11}^{L}}{b_{12}^{L/R}}$, ${a_{12}^{L}}{b_{11}^{L/R}}$, ${a_{21}^{L}}{b_{22}^{L/R}}$, ${a_{22}^{L}}{b_{21}^{L/R}}$\\

$|\psi^-_S\rangle_{AB}|\phi^-_P\rangle_{AB}|\psi^{\pm}_T\rangle_{AB}$&
&${a_{11}^{R}}{b_{22}^{R/L}}$, ${a_{12}^{R}}{b_{21}^{R/L}}$, ${a_{21}^{R}}{b_{12}^{R/L}}$, ${a_{22}^{R}}{b_{11}^{R/L}}$, ${a_{11}^{L}}{b_{22}^{L/R}}$, ${a_{12}^{L}}{b_{21}^{L/R}}$, ${a_{21}^{L}}{b_{12}^{L/R}}$, ${a_{22}^{L}}{b_{11}^{L/R}}$\\

\noalign{\smallskip}\hline
\end{tabular}\label{table1}
\end{table}

\section{Discussion and Conclusion}

Hyper-BSA plays a central role in quantum communication and quantum computation, and numerical theoretical and experimental programs have been proposed.
Complete and deterministic two-photon BSA is not possible by using linear optics alone.\upcite{not-possible}
Bell states can be nondestructively and completely distinguished from each other assisted by cross-Kerr nonlinearity, while giant Kerr nonlinearity is a challenge in experiment.\upcite{kerr-impossible}
Complete neutral-atom-based BSA scheme has been achieved with long coherence time, while individual manipulation and measurement of neutral atoms in optical lattices is not possible.
Additional photons usually are necessary in some complete BSA schemes.\upcite{multiple-DOFs7,guoguangcan,multiple-DOFs3,teleportation-based,multiple-DOFs2,ptd}

\begin{figure} [tpb]
\begin{center}
\includegraphics[width=8 cm,angle=0]{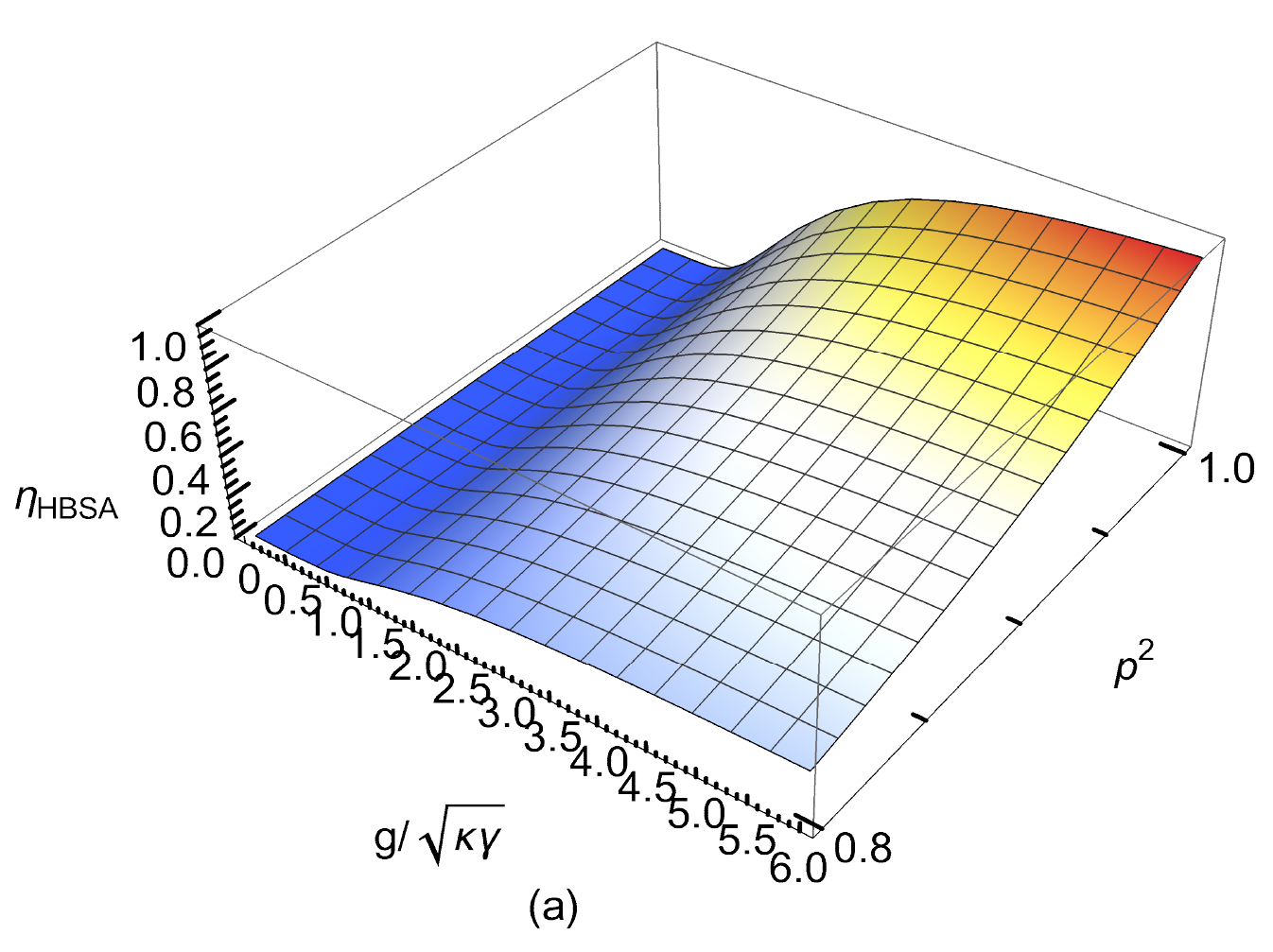}
\includegraphics[width=8 cm,angle=0]{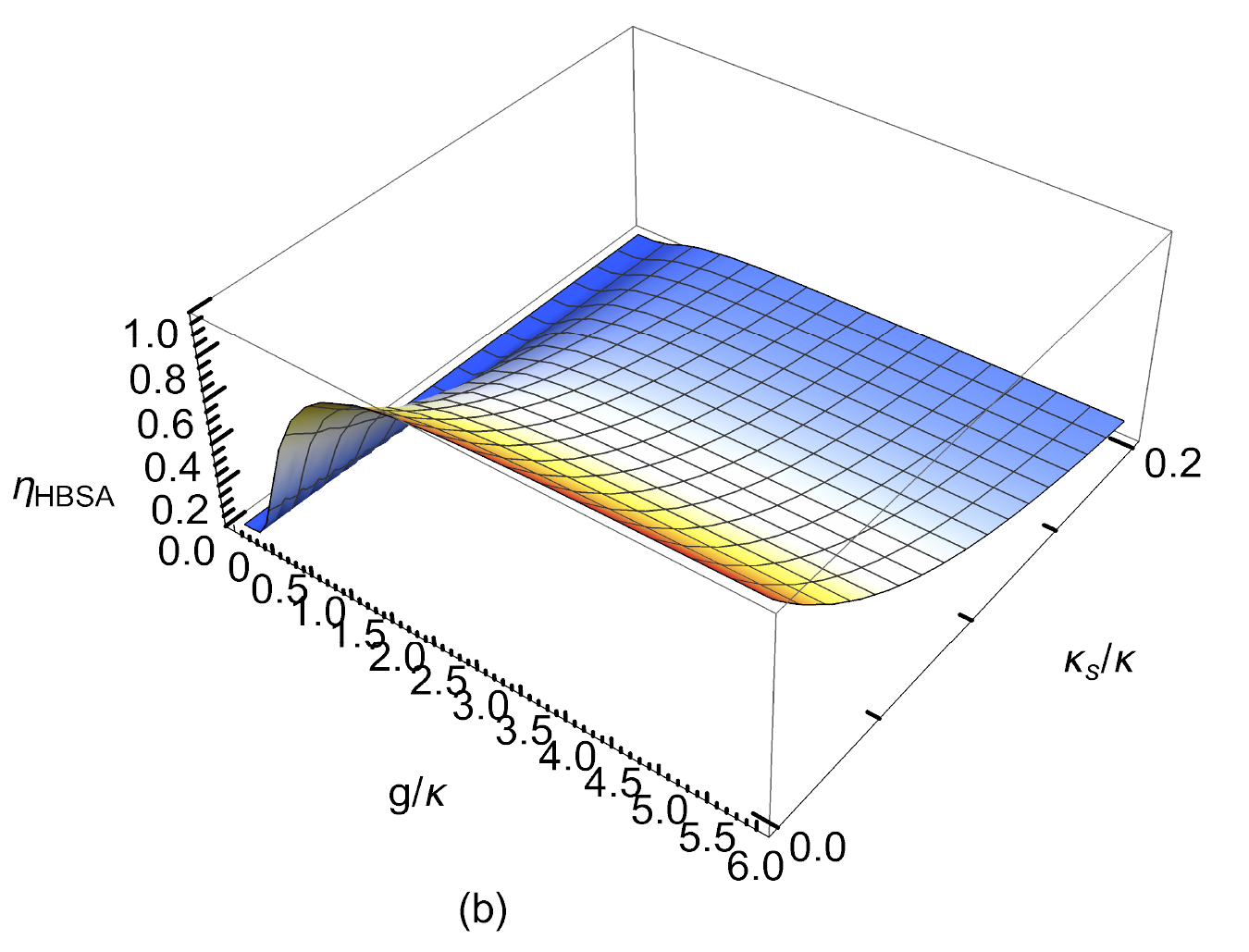}
\caption{(a) The efficiency of the scheme of HBSA using the QD-cavity blocks with parameters $ \kappa_s \ll \kappa$\upcite{sideleakage}
and $\omega_{c}=\omega_{X^-}=\omega$.
(b) The efficiency of the scheme of HBSA using the unity QD-cavity blocks with parameters $\gamma=0.1\kappa$, $\omega=\omega_c=\omega_{X^-}$ and $p=1$.
\label{efficiency}}
\end{center}
\end{figure}

We present a scheme to completely distinguish 64 hyper-Bell states in spatial, polarization, and time-bin DOFs assisted by QD-cavity systems. In the present scheme, the time interval $\Delta t$ between the two incident photons and the cavity photon lifetime ($\tau\sim$ tens of picoseconds)\upcite{cavity-photon1,cavity-photon2} is much shorter than the spin coherence time in QD ($T_2^e\sim \mu$s).\upcite{coherence-QD} Manipulation and readout of the electron spin in QD has been demonstrated with high precision.\upcite{cavity-photon1,manipulation1}
Unity fidelity of our scheme can be achieved with efficiency
$\eta_{\text{HBSA}}=(\frac{p}{2}(r_h-r_0))^{12}$.  Here the fidelity and the efficiency are defined as $F=|\langle \varphi_f|\varphi_i\rangle|^2$ and $\eta= n_{\text{out}}/ n_{\text{in}}$, respectively. $|\varphi_f\rangle$ ($|\varphi_i\rangle$) is the realism (idea) normal output  states of the system. $n_{\text{in}}$ ($n_{\text{out}}$) is the number of the input (output) photons. From Figure \ref{efficiency}, one can see that the high efficiency of our scheme can be achieved by optimizing the QD-cavity system, e.g., increasing the QD-cavity coupling strength and suppressing the cavity side leakage.



The unity fidelity of our scheme can be reduced by few percents due to spin decoherence, trion dephasing including the optical dephasing and the spin dephasing of $X^-$, and the imperfect optical selection rules due to the heavy-light hole mixing.
In addition, several experimental imperfections contribute to reducing the fidelity, including mixing of the photon polarization (0.5\%); background  counts and dark counts (1.2\%); imperfections in the PBS (3\%); spatial mode mismatch between cavity and incident photon (3\%); imperfections in spin state preparation, manipulation, readout, and coherence; more than one photon events (1.2\%).

In summary, we proposed a scheme to complete two-photon hyper-Bell states analysis in polarization, spatial, and time-bin DOFs with three QD-cavity systems, eight PCs and some linear optics. The three QD spins in microcavities are employed as mediate, and strong couple limitation between QD-cavity system is not suffered. Unity fidelity of our scheme can be achieved in principle, the incomplete and imperfect photon-QD interactions are heralded by the single-photon detectors. The protocol reduces the number of the matter medium, and additional photons or DOFs are not required.

\bigskip

\section*{ACKNOWLEDGMENTS}
This work is supported by the National Natural Science Foundation of China under Grant No. 11604012, the Fundamental Research Funds for the Central Universities under Grants FRF-TP-19-011A3, the Natural Science Foundation of China under Contract 61901420; the Shanxi Province Science Foundation for Youths under Contract 201901D211235; the Scientific and Technological Innovation Programs of Higher Education Institutions in Shanxi under Contract 2019L0507, and a grant from the China Scholarship Council.




\begin{thebibliography}{0}
\expandafter\ifx\csname natexlab\endcsname\relax\def\natexlab#1{#1}\fi
\expandafter\ifx\csname bibnamefont\endcsname\relax
  \def\bibnamefont#1{#1}\fi
\expandafter\ifx\csname bibfnamefont\endcsname\relax
  \def\bibfnamefont#1{#1}\fi
\expandafter\ifx\csname citenamefont\endcsname\relax
  \def\citenamefont#1{#1}\fi
\expandafter\ifx\csname url\endcsname\relax
  \def\url#1{\texttt{#1}}\fi
\expandafter\ifx\csname urlprefix\endcsname\relax\def\urlprefix{URL }\fi
\providecommand{\bibinfo}[2]{#2}
\providecommand{\eprint}[2][]{\url{#2}}

\end{thebibliography}


\begin{thebibliography}{}
%
\bibitem{CITE1}M. A. Nielsen, I. L. Chuang, \emph{Quantum Computation and Quantum Information}, Cambridge University Press, Cambridge, \textbf{2000}.

\bibitem{distribution1} A. K. Ekert, \emph{Phys. Rev. Lett.} \textbf{1991}, \emph{67}, 661-663.

\bibitem{distribution2} G. L. Long, X. S. Liu, \emph{Phys. Rev. A} \textbf{2002}, \emph{65}, 032302.

\bibitem{quantum-key-distribution1}  C. C. W. Lim, F. Xu, J. W. Pan, A. Ekert, \emph{Phys. Rev. Lett.} \textbf{2021}, \emph{126}, 100501.

\bibitem{dense1} C. H. Bennett, S. J. Wiesner, \emph{Phys. Rev. Lett}. \textbf{1992}, \emph{69}, 2881-2884.

\bibitem{dense2}  X. S. Liu,  G. L. Long, D. M. Tong, F. Li, \emph{Phys. Rev. A} \textbf{2002}, \emph{65}, 022304.

\bibitem{Advances-in-Quantum-Dense-Coding} Y. Guo, B. H. Liu, C. F. Li, G. C. Guo, \emph{Adv. Quantum Technol.} \textbf{2019}, \emph{2}, 1900011.

\bibitem{teleportation} C. H. Bennett, G. Brassard, C. Cr\'{e}peau,  R. Jozsa,  A. Peres,  W. K. Wootters, \emph{Phys. Rev. Lett.} \textbf{1993}, \emph{70}, 1895-1899.

\bibitem{teleportation1}S. Liu, Y. Lou, J. Jing, \emph{Nat. Commun.} \textbf{2020}, \emph{11}, 3875.

\bibitem{teleportation2}S. Langenfeld, S. Welte, L. Hartung, S. Daiss, P. Thomas, O. Morin, E. Distante, G. Rempe, \emph{Phys. Rev. Lett.} \textbf{2021}, \emph{126}, 130502.


\bibitem{share} M. Hillery, V. Bu\u{z}ek,  A. Berthiaume, \emph{Phys. Rev. A} \textbf{1999}, \emph{59}, 1829-1834.


%
\bibitem{share2} M. D. Oliveira, I. Nape,  J. Pinnell, N. TabeBordbar, A. Forbes, \emph{Phys. Rev. A} \textbf{2020}, \emph{101}, 042303.

\bibitem{direct1}F. G. Deng, G. L. Long, X. S. Liu, \emph{Phys. Rev. A} \textbf{2003}, \emph{68}, 042317.

\bibitem{direct2} W. Zhang, D. S. Ding, Y. B. Sheng, L. Zhou, B. S. Shi, G. C. Guo, \emph{Phys. Rev. Lett.} \textbf{2017}, \emph{118}, 220501.

\bibitem{direct3}  T. Li, G. L. Long, \emph{New J. Phys.} \textbf{2020}, \emph{22}, 063017.

\bibitem{direct4} Z. Zhou, Y. B. Sheng, P. H. Niu, L. G. Yin, G. L. Long, L. Hanzo, \emph{Sci. China Phys. Mech. Astron.} \textbf{2020}, \emph{63}, 230362.

\bibitem{direct5} G. L. Long, H. Zhang, \emph{Sci. Bull. (Beijing)} \textbf{2021}, \emph{66}, 1267.

\bibitem{quantum-router} C. Cao, Y. W. Duan, X. Chen, R. Zhang, T. J. Wang, C. Wang, \emph{Opt. Express} \textbf{2017}, \emph{25}, 16931-16946.

\bibitem{one-way1} R. Raussendorf,  H. J. Briegel, \emph{Phys. Rev. Lett.} \textbf{2001}, \emph{86}, 5188-5191.

\bibitem{one-way2} R. Raussendorf, D. E. Browne, H. J. Briegel, \emph{Phys. Rev. A} \textbf{2003}, \emph{68}, 022312.


\bibitem{multiple-DOFs4} G. Vallone, R. Ceccarelli, F. D. Martini, P. Mataloni, \emph{Phys. Rev. A}  \textbf{2009}, \emph{79}, 030301(R).


\bibitem{multiple-DOFs5} J. T. Barreiro, N. K. Langford, N. A. Peters, P. G. Kwiat, \emph{Phys. Rev. Lett.} \textbf{2005}, \emph{95}, 260501.


\bibitem{multiple-DOFs6} M. Barbieri, C. Cinelli, P. Mataloni, F. D. Martini, \emph{Phys. Rev. A} \textbf{2005}, \emph{72}, 052110.

\bibitem{multiple-DOFs2} C. Schuck, G. Huber, C. Kurtsiefer, H. Weinfurter, \emph{Phys. Rev. Lett.} \textbf{2006}, \emph{96}, 190501.

\bibitem{multiple-DOFs3} M. Barbieri, G. Vallone, P. Mataloni, F. D. Martini, \emph{Phys. Rev. A} \textbf{2007}, \emph{75}, 042317.

\bibitem{multiple-DOFs7} J. T. Barreiro, T. C. Wei, P. G. Kwiat, \emph{Nat. Phys.} \textbf{2008}, \emph{4}, 282-286.


\bibitem{multiple-DOFs1} A. Yabushita, T. Kobayashi, \emph{Phys. Rev. A} \textbf{2004}, \emph{69}, 013806.


\bibitem{teleportation-based} S. P. Walborn, S. P\'{a}dua, C. H. Monken, \emph{Phys. Rev. A} \textbf{2003}, \emph{68}, 042313.


\bibitem{purification1} Y. B. Sheng, F. G. Deng, \emph{Phys. Rev. A} \textbf{2010}, \emph{81}, 032307.

\bibitem{purification2} Y. B. Sheng, F. G. Deng, \emph{Phys. Rev. A} \textbf{2010}, \emph{82}, 044305.

\bibitem{purification3} C. Cao, C. Wang, L. Y. He, R. Zhang, \emph{Opt. Express} \textbf{2013}, \emph{21}, 4093-4105.


\bibitem{entanglement-witness} V. Tr\'{a}vn\'{i}\u{c}ek, K. Bartkiewicz, A. \u{C}ernoch, K. Lemr, \emph{Phys. Rev. A} \textbf{2018}, \emph{98}, 032307.

\bibitem{one-way-qunatum-computing} K. Chen, C. M. Li, Q. Zhang, Y. A. Chen, A. Goebel, S. Chen, A. Mair, J. W. Pan, \emph{Phys. Rev. Lett.} \textbf{2007}, \emph{99}, 120503.



\bibitem{QKD-hyper} D. S. Simon, A. V. Sergienko,\emph{ New J. Phys.} \textbf{2014}, \emph{16}, 063052.

\bibitem{Linear-optical-heralded-amplification}  G. Yang, Y. S. Zhang, Z. R. Yang, L. Zhou, Y. B. Sheng, \emph{Quantum Inf. Process}. \textbf{2019}, \emph{18}, 317.


\bibitem{guoguangcan} X. F. Ren, P. G. Guo, G. C. Guo, \emph{Phys. Lett. A} \textbf{2005}, \emph{343}, 8-11.



\bibitem{ptd} B. P. Williams, R. J. Sadlier, T. S. Humble, \emph{Phys. Rev. Lett.} \textbf{2017}, \emph{118}, 050501.

\bibitem{BSA-kerr} S. D. Barrett, P. Kok, K. Nemoto, R. G. Beausoleil, W. J. Munro, T. P. Spiller, \emph{Phys. Rev. A} \textbf{2005}, \emph{71}, 060302.


\bibitem{NMR} M. A. Nielsen, E. Knill, R. Laflamme, \emph{Nature}  \textbf{1998}, \emph{396}, 52-55.

\bibitem{atom} M. D. Barrett, J. Chiaverini, T. Schaetz, J. Britton, W. M. Itano, J. D. Jost, E. Knill, C. Langer, D. Leibfried, R. Ozeri, D. J. Wineland, \emph{Nature} \textbf{2004}, \emph{429}, 737-739.

\bibitem{QED} L. Ye, G. C. Guo, \emph{Phys. Rev. A} \textbf{2004}, \emph{70}, 054303.


\bibitem{BSA-QD} C. Bonato, F. Haupt, S. S. R. Oemrawsingh, J. Gudat, D. Ding, M. P. van Exter, D. Bouwmeester, \emph{Phys. Rev. Lett.} \textbf{2010}, \emph{104}, 160503.

\bibitem{BSA-QDs} B. C. Ren, H. R. Wei, M. Hua, T. Li, F. G. Deng, \emph{Eur. Phys. J. D} \textbf{2013}, \emph{67}, 30.


\bibitem{hu-QD} C. Y. Hu, J. G. Rarity, \emph{Phys. Rev. B} \textbf{2011}, \emph{83}, 115303.

\bibitem{Cao-QD} L. Fan, C. Cao, \emph{J. Opt. Soc. Am. B} \textbf{2021}, \emph{38}, 1593-1603.



\bibitem{not-possible}  N. L\"{u}tkenhaus, J. Calsamiglia, K. A. Suominen, \emph{Phys. Rev. A} \textbf{1999}, \emph{59}, 3295-3300.

\bibitem{single-photon-BSA} X. X. Chen, J. Z. Yang, X. D. Chai, A. N. Zhang, \emph{Phys. Rev. A} \textbf{2019}, \emph{100}, 042302.

\bibitem{seven-groups} T. C. Wei, J. T. Barreiro, P. G. Kwiat, \emph{Phys. Rev. A} \textbf{2007}, \emph{75}, 060305(R).


\bibitem{sheng} Y. B. Sheng, F. G. Deng, G. L. Long, \emph{Phys. Rev. A} \textbf{2010}, \emph{82}, 032318.

\bibitem{polarization-time-bin} X. H. Li, S. Ghose, \emph{Opt. Express} \textbf{2016}, \emph{24}, 18388.

\bibitem{liu-two-photon-six-qubit} Q. Liu, G. Y. Wang, M. Zhang, F. G. Deng, \emph{Sci. Rep.} \textbf{2016}, \emph{6}, 22016.

\bibitem{three-DOF-kerr-wangmeiyu} M. Y. Wang, F. L. Yan, T. Gao, \emph{Laser Phys. Lett.} \textbf{2018}, \emph{15}, 125206.

\bibitem{three-DOF-kerr-Zhang}  H. R. Zhang, P. Wang, C. Q. Yu, B. C. Ren, \emph{Chin. Phys. B} \textbf{2021}, \emph{30}, 030304.




\bibitem{Self-assisted} X. H. Li, S. Ghose, \emph{Phys. Rev. A} \textbf{2016}, \emph{93}, 022302.


\bibitem{ren-QD} B. C. Ren, H. R. Wei, M. Hua, T. Li, F. G. Deng, \emph{Opt. Express} \textbf{2012}, \emph{20}, 24664-24677.

\bibitem{wang-QD} T. J. Wang, Y. Lu, G. L. Long, \emph{Phys. Rev. A} \textbf{2012}, \emph{86}, 042337.

\bibitem{wang-error-detected-QD}  G. Y. Wang, Q. Ai, B. C. Ren, T. Li, F. G. Deng, \emph{Opt. Express} \textbf{2016}, \emph{24}, 28444-28458.

\bibitem{Error-heralded} Y. Y. Zheng, L. X. Liang, M. Zhang, \emph{Sci. China-Phys. Mech. Astron.} \textbf{2019}, \emph{62}, 970312.

\bibitem{Cao} C. Cao, L. Zhang, Y. H. Han, P. P. Yin, L. Fan, Y. W. Duan, R. Zhang, \emph{Opt. Express} \textbf{2020}, \emph{28}, 2857-2872.

\bibitem{auxiliary-entanglement} X. H. Li, S. Ghose, \emph{Phys. Rev. A} \textbf{2017}, \emph{96}, 020303(R).

\bibitem{time-bin1} C. Y. Gao, B. C. Ren, Y. X. Zhang, Q. Ai, F. G. Deng, \emph{Ann. Phys. (Berlin)} \textbf{2019}, \emph{531}, 1900201.

\bibitem{time-bin2} C. Y. Gao, B. C. Ren, Y. X. Zhang, Q. Ai, F. G. Deng, \emph{Appl. Phys. Express} \textbf{2020}, \emph{13}, 027004.


\bibitem{QD-cavity-gate-hybrid1} H. R. Wei, F. G. Deng, \emph{Phys. Rev. A} \textbf{2013}, \emph{87}, 022305.

\bibitem{QD-cavity-gate-hybrid2} Y. H. Han, C. Cao, L. Fan, Ling Fan, R. Zhang, \emph{Opt. Express} \textbf{2021}, \emph{29}, 20045-20062.

\bibitem{QD-cavity-gate-hybrid3} Y. H. Han, C. Cao, L. Zhang, X. Yi, P. P. Yin, L. Fan, R. Zhang, \emph{Int. J. Theor. Phys.} \textbf{2021}, \emph{60}, 1136-1149.


\bibitem{QD-cavity-gate-solid1} W. Q. Liu, H. R. Wei,  \emph{New J. Phys.} \textbf{2019}, \emph{21}, 103018.

\bibitem{QD-cavity-gate-solid2} C. Cao, Y. H. Han, L. Zhang, L. Fan, Y. W. Duan, R. Zhang, \emph{Adv. Quantum Technol.} \textbf{2019}, \emph{2}, 1900081.



\bibitem{QD-cavity-gate-photon1} B. Y. Xia, C. Cao, Y. H. Han, R. Zhang, \emph{Laser Phys. Lett.} \textbf{2018}, \emph{28}, 095201.


\bibitem{QD-cavity-gate-photon2} H. R. Wei, Y. B. Zheng, M. Hua, G. F. Xu, \emph{Appl. Phys. Express} \textbf{2020}, \emph{13}, 082007.



\bibitem{QD-nature} W. B. Gao, P. Fallahi, E. Togan, J. Miguel-Sanchez, A. Imamoglu, \emph{Nature} \textbf{2012}, \emph{491}, 426-430.

\bibitem{QD-spin} A. Delteil,  Z. Sun, W. B. Gao, E. Togan, S. Faelt,  A. Imamo\u{g}lu, \emph{Nat. Phys.} \textbf{2016}, \emph{12}, 218-223.


\bibitem{teleportation3} W. B. Gao, P. Fallahi, E. Togan, A. Delteil, Y. S. Chin, J. Miguel-Sanchez, A. Imamo\v{g}lu, \emph{Nat. Commun.} \textbf{2013}, \emph{4}, 2744.


\bibitem{QD-NC43} A. J. Bennett, J. P. Lee, D. J. P. Ellis, I. Farrer, D. A. Ritchie, A. J. Shields, \emph{Nat. Nanotechnol.} \textbf{2016}, \emph{11}, 857-860.

\bibitem{QD-NC44} A. Javadi, I. S\"{o}llner, M. Arcari, S. L. Hansen, L. Midolo, S. Mahmoodian, G. Kir\v{s}ansk\.{e}, T. Pregnolato, E. H. Lee, J. D. Song, S. Stobbe, P. Lodahl, \emph{Nat. Commun.} \textbf{2015}, \emph{6}, 8655.


\bibitem{QD-switch} S. Sun, H. Kim, G. S. Solomon, E. Waks, \emph{Nat. Nanotechnol.} \textbf{2016}, \emph{11}, 539-544.


\bibitem{QD-rotation1} C. Arnold, J. Demory, V. Loo, A. Lema\^{i}tre, I. Sagnes, M. Glazov, O. Krebs, P. Voisin, P. Senellart, L. Lanco, \emph{Nat. Commun.} \textbf{2015}, \emph{6}, 6236.

\bibitem{QD-rotation2} P. Androvitsaneas, A. B. Young, C. Schneider, S. Maier, M. Kamp, S. H\"{o}fling, S. Knauer, E. Harbord, C. Y. Hu, J. G. Rarity, R. Oulton, \emph{Phys. Rev. B} \textbf{2016}, \emph{93}, 241409(R).


\bibitem{QD-entangle-photon} Y. H. Han, C. Cao, L. Zhang, X. Yi, P. P. Yin, L. Fan, R. Zhang, \emph{Int. J. Theor. Phys.} \textbf{2020}, \emph{59}, 4025-4039.






\bibitem{QD-basic} C. Y. Hu, A. Young, J. L. O'Brien, W. J. Munro, J. G. Rarity, \emph{Phys. Rev. B} \textbf{2008}, \emph{78}, 085307.

\bibitem{exciton} R. J. Warburton, C. S. D\"{u}rr, K. Karrai, J. P. Kotthaus, G. Medeiros-Ribeiro, P. M. Petroff, \emph{Phys. Rev. Lett.} \textbf{1997}, \emph{79}, 5282-5285.

\bibitem{QD-Pauli-principle} C. Y. Hu, W. Ossau, D. R. Yakovlev, G. Landwehr, T. Wojtowicz, G. Karczewski, J. Kossut, \emph{Phys. Rev. B} \textbf{1998}, \emph{58}, R1766-R1769.

\bibitem{Heisenberg1} D. F. Walls, G. J. Milburn, \emph{Quantum Optics}, Springer-Verlag, Berlin, \textbf{1994}.

\bibitem{Heisenberg2} P. R. Berman, \emph{Cavity quantum electrodynamics}, Academic Press, San Diego, \textbf{1994}.


\bibitem{PC} J. Zhang, X. Zhang, D. Wu, X. Tian, M. Li, F. Jing, \emph{Opt. Express} \textbf{2010}, \emph{18}, A185-A191.



\bibitem{kerr-impossible} J. Gea-Banacloche, \emph{Phys. Rev. A} \textbf{2010}, \emph{81}, 043823.

\bibitem{cavity-photon1} B. D. Gerardot, D. Brunner, P. A. Dalgarno,  P. \"{O}hberg, S. Seidl, M. Kroner, K. Karrai, N. G. Stoltz, P. M. Petroff, R. J. Warburton, \emph{Nature} \textbf{2008}, \emph{451}, 441-444.

\bibitem{cavity-photon2} D. Brunner, B. D. Gerardot, P. A. Dalgarno, G. W\"{u}st, K. Karrai, N. G. Stoltz, P. M. Petroff, R. J. Warburton, \emph{Science} \textbf{2009}, \emph{325}, 70-72.

\bibitem{coherence-QD} H. Bluhm, S. Foletti, I. Neder, M. Rudner, D. Mahalu, V. Umansky, A. Yacoby, \emph{Nat. Phys.} \textbf{2011}, \emph{7}, 109-113.

\bibitem{manipulation1} K. C. Nowack, F. H. L. Koppens, Y. V. Nazarov, L. M. K. Vandersypen, \emph{Science} \textbf{2007}, \emph{318}, 1430-1433.

\bibitem{sideleakage} S. Reitzensteina, C. Hofmann, A. Gorbunovb, M. Strau{\ss}, S. H. Kwon, C. Schneider, A. L\"{o}ffler, S. H\"{o}fling, M. Kamp, A. Forchel, \emph{Appl. Phys. Lett.} \textbf{2007}, \emph{90}, 251109.


\end{thebibliography}
\end{document}